\documentclass[11pt,a4paper]{article}
\usepackage{amsmath}
\usepackage{amssymb}
\usepackage{graphicx}
\usepackage{subcaption}
\usepackage{bbm}
\usepackage{bm}
\usepackage{breqn}
\usepackage{enumitem}
\usepackage[compat=1.1.0]{tikz-feynman}

\newcommand{\beq}{\begin{equation}}
\newcommand{\eeq}{\end{equation}}
\newcommand{\bea}{\begin{eqnarray}}
\newcommand{\eea}{\end{eqnarray}}                  
\newcommand{\bal}{\begin{align}}
\newcommand{\eal}{\end{align}}
\newcommand{\bpm}{\begin{pmatrix}}
\newcommand{\epm}{\end{pmatrix}}
\newcommand{\Ds}{\,{\slash \kern -8pt D}}
\tikzfeynmanset{warn luatex=false}

\makeatletter
\let\cat@comma@active\@empty
\makeatother

\begin{document}

\title{\textbf{Fixed points of the renormalisation group running of quark and fermion mixing  matrices in the Standard Model  and beyond}}

\author{Brian P. Dolan\footnote{email: bdolan@stp.dias.ie} \\
  \textit{\small School of Theoretical  Physics}\\
  \textit{\small Dublin Institute for Advanced Studies}\\
  \textit{\small 10 Burlington Rd., Dublin, Ireland}\\
  and \\
\textit{\small Department of Physics, Maynooth University}\\
  \textit{\small Maynooth, Co.~Kildare, Ireland}\\
}

\maketitle

\begin{abstract}  
    The renormalisation group running of fermion mixing matrices in the Standard model and beyond is studied.  For the massless 1-loop running with three generations six  fixed points are found.   Their associated anomalous dimension matrices are calculated and the nature of each fixed point, whether attractive, repulsive or mixed, is determined.     An argument  is given that the fixed points found at 1-loop must remain fixed points   to all  orders in perturbation theory and even non-perturbatively, as they are associated with certain  differential geometric properties of vector fields on the space of mixing matrices.   With $N_g$ dark or sterile neutrinos there are at least  $N_g!$ fixed points of the fermion mixing matrix.
  \end{abstract}


\section{Introduction}

A central concept in understanding the Standard Model of particle physics, and relating experimental results
to the 19 fundamental parameters in the Standard Model  Lagrangian, is the running of the parameters as a function of energy: the $\beta$-functions.  Understanding the behaviour of $\beta$-functions is also essential in any attempt to go beyond the Standard Model.  The running of the gauge couplings and the discovery of asymptotic freedom \cite{Politzer,Gross+Wilczek} was a crucial step in the development of the Standard Model and the running of the Higgs quartic coupling and the top quark Yukawa coupling is important for the stability of the Higgs sector at high energies \cite{Stability,Shaposhnikov}.

The running of the parameters in the Cabibbo-Kobayashi-Maskawa 
(CKM) matrix \cite{CKM} is not so well known perhaps for two reasons: the measured magnitude  of the quark Yukawa couplings is too small for the running of the CKM parameters to have any physical relevance; and, even at 1-loop, the $\beta$-functions are not simple.  The running of the Pontecorvo-Maki-Nakagawa-Sakata (PMNS) matrix \cite{PMNS,MNS} is more relevant beyond the Standard Model, 
and the mixing matrices in both the quark and the lepton sectors can be treated in parallel with the same techniques:  for the Standard Model with the addition of three gauge singlet, Weyl, right-handed neutrinos  the analysis is identical to that of the quark sector.

The $\beta$-functions for the CKM parameters were investigated in \cite{M+P,  B+G},
in the limit where the top Yukawa coupling $y_t$ dominates, and it was observed there there is an infra-red fixed point.  The full 1-loop $\beta$-functions for the CKM matrix  were calculated in \cite{Babu}.    The 1-loop $\beta$-functions for mixing matrices were further analysed  in  \cite{Denner+Sack}-\cite{Zhang1} but the full expressions  are somewhat complicated, and they are significantly simpler in the small angle approximation:  approximate 1-loop $\beta$-functions, with one dominant Yukawa coupling and/or  at least one  small mixing angle, were analysed in \cite{AKLR} and \cite{GIRT}
(for higher loops, see \cite{3-loops}-\cite{AEHNPS}).  However these approximations obscure an analytic pattern in the
$\beta$-functions which shall be explored in the present paper.

We shall therefore study the analytic form of the full 1-loop massless renormalisation group (RG) equations for the mixing matrix: fixed points are determined and the matrix of anomalous dimensions
is calculated  at each of the fixed points (full 2-loop $\beta$-functions are given
in \cite{M+V, Luo+Xiao}, but we leave this for later analysis).  The $\beta$-functions are of course scheme dependent, but the existence of fixed points,  and the eigenvalues of the matrix of anomalous dimensions at the fixed points, are scheme independent.
 There are 6 fixed points\footnote{After this work was complete we became aware of reference \cite{AEHNPS} where these six fixed points were also found.} 
and the associated mixing matrices form a unitary representation of a group of order  6, the group of permutations of 3 objects, $S_3$.   
The Jarlskog invariant \cite{Jarlskog} vanishes at all
of the fixed points.

The parameters in the mixing matrices consist of three angles and one phase, parameterising a space that is topologically the double coset
\[U(1)\times U(1)\backslash  SU(3)/ U(1) \times U(1),\]
where $U(1)\times U(1)\approx T$ is the Cartan torus of $SU(3)$ (generated by $\lambda_3$ and $\lambda_8$ in the Gell-Mann representation).  The right-coset $SU(3)/U(1) \times U(1)$ is the flag manifold  $F_3$, a compact complex manifold that admits a metric with isometry group $SU(3)$.  
The left action of $T$ on $F_3$ has 6 fixed points. and these are precisely the 6 fixed points of the 1-loop
RG running.

If the RG running is lifted to $F_3$, which can be done in a well defined manner, and it is assumed that the left action of $T$ on   $F_3$ commutes with this running, then a proof is given\footnote{I thank Charles Nash for pointing out the significance of the fact that these actions commute for the  fixed points.}  that fixed points of the left action of $T$  on   $F_3$ must
necessarily be fixed points of the RG.   With this assumption  the fixed points that are found at 1-loop must remain fixed points at all orders in perturbation theory, and even non-perturbatively,

The 1-loop running is reviewed  in \S\ref{sec:Yukawa} and a technique for extracting the running of the mixing parameters from the running of the Yukawa matrices is presented in \S\ref{sec:RG}.  As a simple example the special case of two generations, the Cabibbo angle, is presented in detail in \S\ref{sec:Cabibbo} before the three generation 1-loop $\beta$ functions for the mixing matrices are presented in \S\ref{sec:CKM-PMNS}.   The eigenvalues of the matrix of anomalous dimensions at each of the fixed points  are determined in \S\ref{sec:AD}, and presented in detail in appendix \ref{app:lambda}.  A proof that fixed points of the left action of $T$ on $F_3$ must be fixed points of the RG is given in \S\ref{sec:allorders}.  The significance of the results and possible future developments  are discussed in \S\ref{sec:conclusions}.
Full expression for the 1-loop $\beta$-functions are given in equations \eqref{eq:theta-dot}, \eqref{eq:delta-dot} and appendix \ref{app:PQ}.

\section{Yukawa couplings \label{sec:Yukawa} }

The Yukawa couplings in the Standard Model are
\beq  {\cal L}_{Yukawa} = - \sum_{\bar a, b =1}^3 \left( Y_{\bar a b} \overline \Psi_L^{\bar a} \Phi_c \Psi^b_R
+ Y'_{\bar a b}\overline \Psi_L^{\bar a}\Phi\Psi^{\prime b}_R\right)
\ - \ \mbox{h.c.}
\eeq
where 
$\Phi=\bpm \phi^+\\  \phi^0 \epm $ is the Higgs doublet;
$\Phi_c=-i \sigma_2\Phi^*$ the conjugate Higgs;
$\Psi_L^a$ are left-handed $SU(2)$ doublets; $\Psi^a_R$ and $\Psi^{\prime a}_R$
right-handed singlets. 
The Yukawa couplings $Y_{\bar a b}$ and   $Y'_{\bar a b}$
are $3\times 3$ complex matrices, with $a,b=1,2,3$ labelling generations.

In the quark sector
\[ \Psi_L = \bpm u_{L} & c_L & t_L\\ d_L & s_L & b_L\epm, \quad
 \Psi_R = (u_R, c_R, t_R )\quad \mbox{and} \quad 
  \Psi'_R = (d_R,  s_R,  b_R ) , 
\]
but the ensuing analysis is equally applicable to the Standard Model with the leptonic sector extended by 3 right-handed gauge-singlet Weyl neutrinos,  in which case\footnote{Majorana neutrinos are not considered here: if  the neutrinos are Majorana the analysis would be more complicated as there are more mixing parameters and extra terms involving Majorana masses, \cite{Schechter+Valle}.}
\[ \Psi_L = \bpm \nu_{e,L} & \nu_{\mu,L} & \nu_{\tau,L}\\ e_L & \mu_L & \tau_L\epm, \quad
  \Psi'_R =( e_R,  \mu_R,  \tau_R)\ \mbox{and} \   \Psi_R =  (\nu_{1,R},  \nu_{2,R},  \nu_{3,R}  ).
\]

There are eighteen parameters in $Y_{\bar a b}$ and $Y'_{\bar a b}$ but, as is well known, these are not all physical.  The Weyl fermions  $\Psi_L$ and $\Psi_R$ can be rotated by two 
different $U(3)$ transformations in generation space, to bring $Y_{\bar a b}$ to real diagonal form, with just three Yukawa couplings; a further eight parameters can be removed from  $Y'_{\bar a b}$ with a further $SU(3)$ transformation on $\Psi'_R$, leaving 10 parameters  in  $Y'_{\bar a b}$.  But we are still free to perform individual $U(1)$ phases transformations on the three generations in $\Psi_L$ (compensating with phases transformation on $\Psi_R$ to keep $Y_{\bar a b}$ real)  to  remove another three parameters from $Y'_{\bar a b}$, leaving seven parameters in $Y'_{\bar a b}$: three real Yukawa couplings and the four mixing parameters of the CKM (or PMNS)  matrix.

The 1-loop running of the Yukawa matrices $Y$ and $Y'$ can be determined by standard techniques; the  Feynman diagrams that contribute  are shown in appendix \ref{app:Feynman}. Using dimensional regularisation, with $t=\ln \mu$, the massless running is  \cite{Babu}, 
\begin{align}
  \frac{d  Y}{d  t} &=
  \frac{1 }{16\pi^2}\left[\frac{3}{2}\left( Y Y^\dagger-   Y'Y^{\prime\dagger}\right)
                                   +4 \pi (\Pi -\alpha){\bm 1}\right]Y,
  \\
   \frac{d  Y'}{d  t} &=
  \frac{1 }{16\pi^2}\left[\frac{3}{2}\left( Y' Y^{\prime \dagger}-   Y Y^\dagger\right)
                       + 4 \pi (\Pi -\alpha'){\bm 1}\right]Y',
                       \end{align}
                       where $\Pi$ arises from summing over fermion loops in the Higgs leg of the Yukawa vertex and $\alpha$ and $\alpha'$ are quadratic combinations of  the  gauge
couplings $\alpha_3$, $\alpha_2$ and $\alpha_1$
(their explicit form will not be relevant for the following analysis).\footnote{Explicitly:
$\Pi =  \frac{1}{4 \pi}\left(3Tr( Y_q Y^\dagger_q+Y'_q Y^{\prime\dagger}_q) +Tr( Y'_l Y^{\prime\dagger}_l) \right)$,
 where $q$ stands for quarks and $l$ for leptons (for $\nu$SM there is a fourth term:  
$ \frac{1}{4 \pi}Tr( Y_l Y^{\dagger}_l)$).
For quarks
$\alpha = 8\alpha_3 +\frac{9}{4}\alpha_2 + \frac{17}{12}\alpha_1$
and
$\alpha'= 8\alpha_3 +\frac{9}{4}\alpha_2 + \frac{5}{12}\alpha_1$,
while for leptons 
$\alpha = \frac{9}{4}\alpha_2 + \frac{3}{4}\alpha_1$ and  $\alpha'=\frac{9}{4}\alpha_2 + \frac{15}{4}\alpha_1$,
$\alpha_i = \frac{g_i^2}{4\pi}$ being the gauge couplings ($\alpha_1$ is hypercharge).}

It is convenient to express the running in terms of Hermitian matrices
\[ Z=\frac{1}{4\pi} Y Y^\dagger \quad \mbox{and}
\quad Z' = \frac{1} {4\pi}  Y' Y^{\prime \dagger},\]
 in terms of which
 \begin{align}
\frac{d  Z}{d  t} &=
\frac{1}{4 \pi} \left\{   3 Z^2-\frac{3}{2}  \left(Z Z' + Z' Z\right)
 +2\left(\Pi -\alpha\right)) Z\right\},\label{eq:dZdt}\\
\frac{d  Z'}{d  t} &=
   \frac{1}{4\pi} \left\{3 Z^{\prime 2}-\frac{3}{2}  \left(Z Z' + Z' Z\right)
                     + 2(\Pi - \alpha') Z'\right\}.\label{eq:dZ'dt}
                     \end{align}

   The Hermitian matrices $Z$ and $Z'$ can be  diagonalized with $U$ and $U'\in SU(3)$:
\beq \Lambda  = U^\dagger Z U,
  \qquad \Lambda ' = U^{\prime \dagger} Z' U'
  \label{eq:ZLambda}\eeq
  where the diagonal components of $\Lambda$ and $\Lambda'$ are related to the usual Yukawa couplings by
 \[\Lambda _a=\frac{y_a^2}{4\pi} ,  \qquad 
    \Lambda '_a= \frac{y_a^{\prime 2}}{4\pi}.\]
  
If  $Z$ and $Z'$ do not commute, then $U\ne U'$ and
$$
V = U^\dagger U'$$
is the mixing matrix, the CKM matrix for quarks and the PMNS matrix for leptons.

If the right-handed neutrinos are Weyl, and not Majorana, the $U(1)$ phases of the leptons, as well as the quarks,  are unobservable and, for both quarks and leptons,  the space of physical parameters in the mixing matrix $V$  is the double coset $U(1)\times U(1)\backslash  F_3$. 
This is a four dimensional space
so there are four physical parameters in $V$.\footnote{In \cite{Buchstaber+Terzic} a proof is given that
$T \backslash F_3$ is topologically $S^4$,  though it is not everywhere differentiable --- in the same way as the surface of a cube, while not a differentiable  manifold,  is topologically $S^2$.
I thank Charles Nash for bringing my attention to  this reference.} 
A general $SU(3)$ matrix can be parameterised in terms of the Gell-Mann matrices as
\[ V= e^{i (\psi_6  \lambda_6 + \psi_7\lambda_7)}
  e^{i (\psi_4 \lambda_4 + \psi_5 \lambda_5)}
  e^{i (\psi_1\lambda_1 + \psi_2\lambda_2)}
  e^{i(\psi_3 \lambda_3 + \psi_8 \lambda_8)}.\]
The phases $\psi_3$ and $\psi_8$ can then be eliminated by the right action of $U(1)\times U(1) $
to render
$ V \in SU(3)/U(1) \times U(1)$.
Changing variables to 

{\small{
\begin{align*}
\psi_1 &=-\theta_{3} \sin\phi_{3}, \  \psi_2=\theta_{3} \cos\phi_{3},\\
      \psi_4 &=\ \ \theta_{2} \sin\phi_{2}, \  \psi_5=\theta_{2} \cos\phi_{2},\\ 
   \psi_6 &=-\theta_{1} \sin\phi_{1}, \  \psi_7=\theta_{1} \cos\phi_{1}
     \end{align*}
   }}

\noindent this is\footnote{The usual physics notation is $\theta_1=\theta_{23}$, $\theta_2 =\theta_{31}$, $\theta_3=\theta_{12}$, which emphasises the generations being mixed: the notation adopted here is chosen to reduce the number of indices.}

{\small{
\beq
  \hskip -10pt V=  \bpm 1 & 0 & 0  \\
 0 & c_{1} & s_{1} e^{i \phi_{1}}  \\
  0 & -s_{1} e^{-i\phi_{1}} & c_{1} \epm
              \kern -4pt
 \bpm c_{2} & 0 & s_{2} e^{-i \phi_{2}}  \\
 0 & 1 & 0  \\
-s_{2} e^{i\phi_{2}} &0&  c_{2}  \epm
 \kern -4pt
                           \bpm c_{3} & s_{3} e^{i \phi_{3}} & 0 \\
-s_{3} e^{-i\phi_{3}} & c_{3} & 0  \\ 0 & 0 & 1\epm \kern -2pt,  \label{eq:V1V2V3} 
\eeq
}}

\noindent
where $s_{i} = \sin\theta_{i}$ and $c_{i} =\cos\theta_{i} $.
The space of all such $V$ is the complex manifold $SU(3)/U(1)\times U(1)\approx F_3$.
This form of $V$ is preserved by the adjoint action of $h\in U(1)\times U(1)$,

{\small{
\[
  h  V h^\dagger =
 \bpm 1 & 0 & 0  \\
 0 & c_{1} & s_{1} e^{i {\tilde \phi}_{1}}  \\
  0 & -s_{1} e^{-i{\tilde \phi}_{1}} & c_{1} \epm
 \bpm c_{2} & 0 & s_{2} e^{-i {\tilde \phi}_{2}}  \\
 0 & 1 & 0  \\
-s_{2} e^{i{\tilde \phi}_{2}} &0&  c_{2}  \epm
                           \bpm c_{3} & s_{3} e^{i {\tilde \phi}_{3}} & 0 \\
-s_{3} e^{-i{\tilde \phi}_{3}} & c_{3} & 0  \\ 0 & 0 & 1\epm,
\]
}}

\noindent
with $\tilde \phi_{1} + \tilde \phi_{2}+\tilde \phi_{3}=  \phi_{1} + \phi_{2} + \phi_{3}$.
The two angles in $h$ can be chosen so that ${\tilde \phi}_{1} = {\tilde \phi}_{3}=0$
putting $ V$ into the standard form, \cite{CKMparamaterisation},
\begin{align}
  V &=\bpm 1 & 0 & 0  \cr
 0 & c_{1} & s_{1}  \cr
0 & -s_{1}  & c_{1} \epm
 \bpm c_{2} & 0 & s_{2} e^{-i\delta} \cr
 0 & 1 & 0  \cr
-s_{2} e^{i\delta} &0&  c_{2}  \epm
 \bpm c_{3} & s_{3} & 0 \cr
-s_{3}  & c_{3} & 0  \cr 0 & 0 & 1 \epm\cr
&=\bpm
c_{3}c_{2} & s_{3}c_{2} & s_{2} e^{-i\delta} \cr
-s_{3}c_{1}-c_{3}s_{1}s_{2}e^{i\delta} 
& c_{3}c_{1} -s_{3}s_{1}s_{2}e^{i\delta} & s_{1}c_{2} \cr 
s_{3}s_{1} - c_{3}c_{1}s_{2}e^{i\delta} &
    -c_{3}s_{1}-s_{3}c_{1}s_{2}e^{i\delta} & c_{1}c_{2}\epm \hskip -2pt ,
      \label{eq:Vdelta}
\end{align}
with ${\tilde \phi}_{2}=\delta$.
This parameterisation  can be used  for both the CKM matrix and for the PMNS matrix with Weyl neutrinos  (strictly speaking it would be $V^\dagger$ that is the standard convention for the PMNS matrix),
but, more generally,  we can  use \eqref{eq:V1V2V3} with $C\kern -2pt P$-violating phase $\delta=\phi_{1}+  \phi_{2} + \phi_{3}$.

The physics is invariant under independent left and right actions of $U(1)\times U(1)$, 
\[  V  \rightarrow  h  V h^{'\dagger},\]
with $h$ acting on $\Psi_R$ and and $h'$ acting on $\Psi'_R$.
Such transformations will be referred to as phase transformations, it being understood that  $h$ and $h'$ are global, but they may depend on the RG scale $\mu$.
We shall also refer to $y_a$ and $y'_a$ as Yukawa couplings and $(\theta_1,\theta_2,\theta_3,\delta)$ as mixing angles and a phase.

\section{Running couplings \label{sec:RG} }
 
Allowing for possible scale dependence $t=\ln\mu$ in $U(\mu)$, $U'(\mu)$,
the RG evolution of equations \eqref{eq:ZLambda} gives
\begin{align}
\frac{d   Z}{ dt}  &=   U
\left(\frac{ d  \Lambda}{dt}  
 - i [A,\Lambda]\right)  U^\dagger, \label{eq:dLambdadt}\\
\frac{ d  Z'}{dt} &=   U'
\left( \frac{d  \Lambda'}{ dt} 
                    - i [A',\Lambda']  \right)   U^{\prime \dagger}, \label{eq:dLambda'dt}
\end{align}
where
\beq
A = i \, U^\dagger \frac{d U}{d t}, \qquad
A' = i \, U^{\prime \dagger}\frac{d \, U'}{d t}. \eeq
In terms of $A$ and $A'$
\beq \frac{d  V}{d t} =i ( A V - V A'  ),\label{eq:Vdot}
\eeq
or 
\[ \frac{d V}{d t} = i B V\]
with
\beq B=-i \frac{d V}{d t} V^\dagger = A - V A' V^\dagger.\label{eq:Bdef} \eeq
  
The off-diagonal components of the two  matrices $A$ and   $A'$ can be obtained from  
equations \eqref{eq:dZdt} and \eqref{eq:dZ'dt}, expressed in terms of the
diagonal matrices  $\Lambda$ and $\Lambda'$:
\begin{align}
  \frac{d \Lambda}{dt}&=
 \frac{1}{4\pi} \left\{3\Lambda^2 -
 \frac{3}{2}(\Lambda V \Lambda' V^\dagger +  V \Lambda' V^\dagger\Lambda )
 +2\bigl( \Pi  - \alpha\bigr) \Lambda  \right\} + i[A,\Lambda],\label{eq:A}\\
 \frac{d\Lambda'}{dt}&=
                      \frac{1}{4 \pi} \left\{3 \Lambda^{\prime 2}
                      - \frac{3}{2}(\Lambda'  V^\dagger \Lambda  V +  V^\dagger \Lambda  V \Lambda')
                      +2 \bigl( \Pi -\alpha'\bigr) \Lambda'  \right\}
                             + i[A',  \Lambda'],\label{eq:A'}
\end{align}
by demanding that the right-hand side of each these equations is diagonal.
 Since $A$ and $A'$ are  hermitian matrices  in the Lie algebra of $SU(3)$
they can both be expanded in terms of Gell-Mann matrices, $A = A_I \lambda_I$, $A'=A'_I \lambda_I$.  However equations \eqref{eq:A} and  \eqref{eq:A'} 
 do not involve the diagonal components $A_3$, $A_8$, $A^'_3$ or $A'_8$;
\eqref{eq:A} and  \eqref{eq:A'}  are sufficient to 
fix $A_1$, $A_2$, $A_4$, $A_5$, $A_6$, $A_7$
and $A'_1$, $A'_2$, $A'_4$, $A'_5$, $A'_6$, $A'_7$ in terms of the off-diagonal components of  the anti-commutators,
\[\{\Lambda, V \Lambda' V^\dagger\} \qquad\mbox{and}
\qquad \{\Lambda' , V^\dagger \Lambda  V\} ,
 \]
 but
$A_3$, $A_8$, $A'_3$ and $A'_8$ remain undetermined.
   A natural choice is to choose fermion phases which preserve the parameterisation
 \eqref{eq:Vdelta} of $ V$ at all energies.
  This requires that the four components
  $\frac{d  V_{11}}{ d t}$, $\frac{d  V_{12}}{d t}$, $\frac{d  V_{23}}{d t}$ and
$\frac{d  V_{33}}{d t}$ of $\frac{d V}{d t}$ must be real and equation \eqref{eq:Vdot} then gives  linear equations for 
$A_3$, $A_8$, $A'_3$ and $A'_8$,
which can  be solved to determine them uniquely as functions of the parameters in $V$
and the off-diagonal components of $A$ and $A'$, which have already been  calculated. 

 An important consequence of this is that $A$ and $A'$ do not depend on $\Pi$, $\alpha$ or $\alpha'$;
they are completely determined by demanding that the off-diagonal components of
\[-\frac{3}{8\pi} \{\Lambda, V \Lambda'  V^\dagger \}  +i[A,\Lambda ]\qquad
\mbox{and} \qquad   
 -\frac{3}{8\pi} \{ V^\dagger \Lambda  V,\Lambda' \}  +i[A',\Lambda']\]
vanish, together with the phase convention that $\dot V_{11}$, $\dot V_{12}$, $\dot V_{23}$ and $\dot V_{33}$ be real.
 In this way $A$ and $A'$ can be expressed  uniquely as functions of the four parameters  in $ V$ and the six Yukawa couplings in $\Lambda$ and $\Lambda'$.  Once $A$ and $A'$ are known the $\beta$-functions for the parameters in $V$ are obtained from \eqref{eq:Vdot}. 

More generally  one can use \eqref{eq:V1V2V3}, with the weaker condition that
only  $\dot V_{11}$ and  $\dot V_{33}$ are real.   This is sufficient to fix
the differences
$A_3 - A'_3$  and $A_8-A'_8$, but leaves the sums $A^+_3=A_3 + A'_3$ and
$A^+_8=A_8 + A'_8$ arbitrary.
Only $\delta=\phi_{1}+ \phi_{2} + \phi_{3}$ is physical,
one is free to choose arbitrary energy dependent phases $\phi_{1}(t)$ and $\phi_{3}(t)$
in order to determine $A^+_3$ and $A^+_8$, but all physical quantities are independent 
of $A^+_3$ and $A^+_8$ and only the
combination $\delta$ remains.  It is useful to perform all calculations in a general such ``gauge'', as a check on our results: $\phi_1$, $\phi_3$, $A_3^+$ and $A_8^+$ will be left arbitrary, but they must drop out in the calculation of any physically quantity, only
$\delta=\phi_1+\phi_2+\phi_3$ should remain.
This is the strategy that will be adopted in \S\ref{sec:CKM-PMNS}, but we first
examine the case of two generations to warm up.
  
\section{Two generations: the Cabibbo angle \label{sec:Cabibbo}}

 Consider first the quark sector in the case of two generations,
 $\bpm u \\ d \epm$ and $ \bpm c \\ s \epm$,
 when the parameters of the CKM matrix reduce to the Cabibbo angle $\theta_C$,
 with $\theta_C \in [0,\frac{\pi}{2}]$, and $ V$ reduces to the $2\times 2$ matrix
 \[ V  = \bpm \cos\theta_C & \sin\theta_C \\ -\sin\theta_C & \cos \theta_C\epm.\]
 Equations \eqref{eq:A} and \eqref{eq:A'} give the off diagonal components of $i[A,\Lambda]$ and $i[A,\Lambda']$ from the off-diagonal components of the anti-commutators
\beq \frac{3}{8\pi}\{V  \Lambda' V ^T ,\Lambda\}\quad
\mbox{and} \quad
\frac{3}{8\pi }\{V ^T \Lambda V ,\Lambda'\}\label{eq:V anticommutators}\eeq
respectively, with
\[ \Lambda= \frac{1}{4\pi}\bpm y_u^2 & 0 \\ 0 & y_c^2\epm
  \quad \mbox{and} \quad
  \Lambda'=  \frac{1}{4\pi}\bpm y_d^2 & 0 \\ 0 & y_s^2\epm.\]
The anti-commutators are purely real and 
\begin{align*}
  A =& \bpm A_3 &-\frac{3 i}{64 \pi^2}\left(\frac{(y_c^2+y_u^2) (y_s^2-y_d^2) }{ y_c^2-y_u^2} \right) \sin 2 \theta_C\\
\frac{3 i}{64 \pi^2}\left(\frac{ (y_c^2+y_u^2) (y_s^2-y_d^2)  }{  y_c^2-y_u^2}  \right)\sin 2 \theta_C & -A_3\epm,\\
  A' = &\bpm A'_3 & -\frac{3 i}{64 \pi^2}\left(\frac{(y_d^2 + y_s^2) (y_c^2 - y_u^2)}{ y_d^2 - y_s^2}\right) \sin 2 \theta_C\\
  \frac{3 i}{64 \pi^2}\left(\frac{ (y_d^2 + y_s^2) (y_c^2 - y_u^2) }{y_d^2 - y_s^2} \right)\sin 2 \theta_C  & -A'_3\epm,
                  \end{align*}
 with $A_3$ and $A_3'$ yet to be determined.
Demanding that $\frac {d V }{d t}$ is real in the 2-generations version of \eqref{eq:Vdot},
\[  \frac{d  V}{d t} =i (  A V- V A' ) ,\]
then requires  $A_3=A'_3=0$, so
\begin{align*}
  A &= \frac{3  }{64 \pi^2 }\left(\frac{(y_c^2+y_u^2) (y_s^2-y_d^2) \sin 2 \theta_C}{ y_c^2-y_u^2}\right)  \sigma_2\\
  A' &= \frac{3  }{64 \pi^2}\left(\frac{(y_d^2+y_s^2) (y_c^2-y_u^2) \sin 2 \theta_C}{ y_d^2-y_s^2}\right)  \sigma_2.
\end{align*}
These expressions  for $A$ and $A'$ then give the Cabibbo angle $\beta$-function for the 2-generation version of
\eqref {eq:Bdef} as
\[  B=A-V  A' V^T -A  = A-A'=\beta_C \sigma_2
  \]
  with  \cite{M+P}
  \[ \beta_C=\frac{3 \sin 2 \theta_C}{64 \pi^2 }
\left[    \left(\frac{y_c^2+ y_u^2 }{y_c^2-y_u^2}\right)(y_s^2-y_d^2)
      +\left(\frac{y_s^2+y_d^2}{y_s^2-y_d^2}\right) (y_c^2-y_u^2)\right] .\]

   It proves convenient to define the variables (similar variables will be used extensively in the next section)
  \[ z = \frac{y_c^2-y_u^2}{y_c^2+y_u^2}, \qquad  z' = \frac{y_s^2-y_d^2}{y_s^2+y_d^2},\]
  which lie in the range $[-1,1]$: observationally these are both close to 1. 
  In terms of $z$ and $z'$
  \beq   \frac{d \theta_C}{d t} =\beta_C=  \frac{3 \sin 2 \theta_C}{32 \pi^2 }  \left[ \frac{y_c^2 z}{z'(1+z)}+\frac{y_s^2 z'}{z(1+z')}
  \right].\label{eq:beta_Cabibbo}
  \eeq
There are fixed points of $\theta_C$ at $\theta_C=0$ and $\frac{\pi}{2}$.
 
Equation \eqref{eq:beta_Cabibbo} is easily solved, with $\theta_C=\theta_0$ at 
$t=\ln\frac{\mu}{\mu_0}=0$,  to give
\beq \tan \theta_C = \tan\theta_0 \,e^{\int b d t}\approx \tan\theta_0 \left(\frac{\mu}{\mu_0}\right)^b \label{eq:thetaC}\eeq
where
\[b= \frac{3 }{16 \pi^2 }  \left[
      \frac{y_c^2 z}{z'(1+z)}+\frac{y_s^2 z'}{z(1+z')}\right].\]
Thus $\theta_C \rightarrow 0$ or $\frac{\pi}{2}$ for $\mu \rightarrow \infty$ (depending on the sign of $b$).\footnote{This analysis is only for illustrative purposes: equation
\eqref{eq:thetaC} is not  valid for energies of the order of, or less than, the Higgs mass, since only massless RG evolution is considered here: the running will freeze  for energies close to, and below, the Higgs mass.}
With the hierarchy $y_u\approx y_d \ll y_s^2 \ll y_c^2$,
\[ \frac{d \theta_C}{d t} \approx  \frac{3  y_c^2 \sin 2 \theta_C} {64 \pi^2}.\]


Note that the poles in $\beta_C$  at $z=0$ and $z'=0$ are not a pathology \cite{CEIN}.
For example, if  the RG trajectories of $y_s$ and $y_d$ cross at some energy $t$, so $y_s^2 = y_d^2$ and $z'=0$ there, then $\Lambda'$ is proportional to the identity matrix
at that energy.   With $y_c^2>y_u^2$ and $y_s^2 - y_d^2 = \epsilon>0$,
$\frac{ d \theta_C}{d t}\rightarrow \infty$  as $\epsilon\rightarrow 0$  and the RG evolution drives $\theta_C$
infinitely quickly to  $\frac{\pi}{2}$  as $\epsilon \rightarrow 0$.
This is  physically perfectly reasonable.

\begin{figure}
{\centering \includegraphics[scale=0.5]{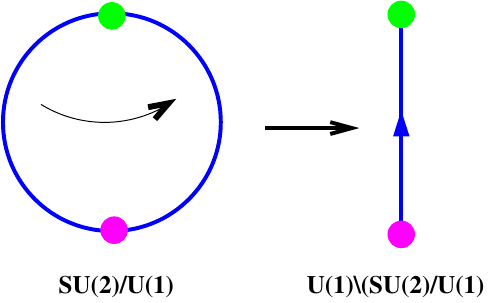}

}
\caption{The left action of $U(1)$ on $SU(2)/U(1)$, with the N and S-poles fixed points, reduces the sphere to a line of
constant longitude.}
\label{fig:SU2}
\end{figure}

Geometrically the Cabibbo angle parameterises  the double coset\newline
$U(1) \backslash SU(2) /U(1)$.  Performing the right action first
$SU(2)/U(1) \approx S^2$, however the left action of $U(1)$ on $S^2$ has fixed points at the N and S-poles.  Using  $(\vartheta,\varphi)$ as standard polar coordinates  on $S^2$, $\vartheta=2 \theta_C$ and  the left action of $U(1)$ is the Killing vector $\frac{\partial}  {d \varphi}$.
 In terms of $\sin\theta_C$, $U(1) \backslash SU(2) /U(1)$ is just the unit line interval $[0,1]$ (see figure \ref{fig:SU2}).

\section{The fermion mixing matrix for three generations \label{sec:CKM-PMNS}}

With three generations, including three right-handed Weyl, gauge singlet,  neutrinos,
the CKM matrix and the PMNS matrix can be treated with the same formalism.
Let $ V$ denote either the CKM matrix in the quark sector or the PMNS matrix in the leptonic sector, so $(y_1, y_2,y_3)=(y_u,y_c,y_t)$,   $(y'_1, y'_2,y'_3)=(y_d,y_s,y_b)$ in the quark sector
and $(y_1', y'_2,y'_3)=(y_e,y_\mu,y_\tau)$,   $(y_1, y_2,y_3)=(y_{\nu_1},y_{\nu_2},y_{\nu_3})$
in the leptonic sector.
To exhibit the explicit form of the $\beta$-functions 
it is convenient to define the ratios
 \begin{align}
   z_1 &=\frac{y_3^2 - y_2^2}{y_3^2+y_2^2}, \qquad
         z_2 =\frac{y_3^2 - y_1^2}{y_3^2+y_1^2},\qquad
          z_3 =\frac{y_2^2 - y_1^2}{y_2^2+y_1^2},\\
   z'_1 &=\frac{y_3^{\prime 2} - y_2^{\prime 2}}{y_3^{\prime 2}+y_2^{\prime 2}}, \qquad
         z'_2 =\frac{y_1^{\prime 2} - y_3^{\prime 2}}{y_1^{\prime 2}+y_3^{\prime 2}},\qquad
          z'_3 =\frac{y_2^{\prime 2} - y_1^{\prime 2}}{y_2^{\prime 2}+y_1^{\prime 2}}.\label{eq:xz-def-q}\end{align}
These all lie in the range $[-1,1]$.  
They are not all independent, of course: for example
        \[ z_3 =\frac{z_2 - z_1}{1-z_1 z_2}.\] 

          In these variables the $\beta$-functions are:
                    \begin{align}
                      \beta_i=\dot \theta_{i} & =\frac{3}{32 \pi^2}
                                     \sum_{j,k=1}^3         \left[
        y_3^2    P^{j k}_{i}  \frac{ z_k }{z'_j(1+z_k)}    
       +     y_3^{\prime 2}     Q^{j k}_{i}   \frac{z'_{k}}{z_j(1+z'_{k}) } 
                                         \right],
                      \label{eq:theta-dot}\\                    
  \beta_\delta=  \dot \delta &=\frac{3 \sin\delta }{32 \pi^2}
       \sum_{j,k=1}^3         \left[
         y_3^2      P^{j k}_\delta  \frac{  z_k }{z'_{j}(1+z_k)} 
         +     y_3^{\prime 2}  Q^{j k}_\delta      \frac{  z'_{k}}{z_j(1+z'_{k}) } 
                               \right],\label{eq:delta-dot}
                    \end{align}
with the $P$s and $Q$s polynomials in the trigonometric functions in $V$.
Explicit expressions for all the $P$ and $Q$ are  given in appendix \ref{app:PQ}.
                                                                 
          There are poles when any of the $z_j$ or  $z'_{j'}$ are zero:  as for the Cabibbo angle above, near a pole the RG flow will push $ V$ very  quickly to a fixed point of the left action of $U(1)\times U(1)$ on the flag manifold $SU(3)/U(1) \times U(1)$.

Using the $P$s and $Q$s in appendix \ref{app:PQ}, the $\beta$-functions \eqref{eq:theta-dot} and \eqref{eq:delta-dot} vanish  when
          \begin{align*}
            \left(\theta^*_{1}, \theta^*_{2}, \theta^*_{3}\right)
= &\left( 0,0,0\right), \
            \left(0,0,\frac{\pi}{2}\right),\
            \left(\frac{\pi}{2},0,0\right),\
                    \left(\frac{\pi}{2},0,\frac{\pi}{2}\right), 
   \\
            &
            \left(0,\frac{\pi}{2},0\right),\
                     \left(\frac{\pi}{2},\frac{\pi}{2},0\right),\
                \left(0,\frac{\pi}{2},\frac{\pi}{2}\right),\
            \left(\frac{\pi}{2},\frac{\pi}{2},\frac{\pi}{2}\right),
          \end{align*}
          with $\delta=0$ or $\pi$.
   For $\theta_{2}=0$ there are four such  points
\[ V^*=
  \begin{pmatrix}
1&0&0\\ 0&1&0\\ 0 &0&1
  \end{pmatrix},\ 
  \begin{pmatrix}
0&1&0\\ -1&0&0\\ 0 &0&1
  \end{pmatrix},\ 
\begin{pmatrix}
1&0&0\\ 0&0&1\\ 0 &-1&0
\end{pmatrix},\
\begin{pmatrix}
0&1&0\\ 0&0&1\\ 1 &0&0
\end{pmatrix},\]
choosing  $\delta=0$ or $\pi$  does not give different  points, $\delta$ only appears in the  mixing matrix $ V$ in the combination $\sin\theta_{2}e^{\pm i\delta}$.

   For $\theta_{2}=\frac{\pi}{2}$ the situation is more subtle.
For $\delta=0$ the four points give:
\[
 V^*=
  \begin{pmatrix}
0&0&1\\ 0&1&0\\ -1 &0&0
  \end{pmatrix},\
 \begin{pmatrix}
0&0&1\\ 0&-1&0\\ 1 &0&0
 \end{pmatrix},\ 
\begin{pmatrix}
  0&0&1\\ -1   &0&0\\ 0 &-1&0
\end{pmatrix},\
\begin{pmatrix}
  0&0&1\\ -1   &0&0\\ 0 &-1&0
\end{pmatrix},
\]
for $\delta=\pi$ they give:
\[
 V^*=
  \begin{pmatrix}
0&0&-1\\ 0&1&0\\ 1 &0&0
  \end{pmatrix},\
 \begin{pmatrix}
0&0&-1\\ 0&1&0\\ 1 &0&0
 \end{pmatrix},\ 
\begin{pmatrix}
  0&0&-1\\ 1   &0&0\\ 0 &-1&0
\end{pmatrix},\
\begin{pmatrix}
  0&0&-1\\ -1   &0&0\\ 0 &1&0
\end{pmatrix}.
\]
Up to phase transformations only two of these eight are distinct physical points.
Furthermore note that, when $\theta_{2}=\frac{\pi}{2}$, the
mixing matrix $ V$ in \eqref{eq:Vdelta} takes the form:
\beq  V=\begin{pmatrix}
0&0&1\\ -\sin \theta_+& \cos\theta_+ & 0\\ -\cos\theta_+ & -\sin\theta_+& 0
\end{pmatrix}\ \mbox{for}\   \delta=0,\label{eq:V20}\eeq 
with $\theta_+=\theta_{3}+\theta_{1}\in[0,\pi]$; and
\beq  V=\begin{pmatrix}
0&0&-1\\ \sin\theta_-& \cos\theta_- & 0\\ \cos\theta_- & -\sin\theta_-& 0
\end{pmatrix}\  \mbox{for}\   \delta=\pi,\label{eq:V2pi}\eeq
with  $\theta_-=\theta_{3}-\theta_{1}\in[-\frac{\pi}{2}, \frac{\pi}{2}]$,
but equations \eqref{eq:V20} and \eqref{eq:V2pi} are physically the same up to phase transformation.
For $\delta=0$ the range 
\[0 \le \theta_+\le  \pi\]
double counts physical values, as the sign of $\cos\theta_+$ in \eqref{eq:V20} can be changed by a phase transformation.  
All distinct physical values of $ V$ with $\theta_{2}=\frac{\pi}{2}$ are achieved with the smaller range
$0\le \theta_+\le \frac{\pi}{2}$.

The upshot of this is that 
 $\theta_{2}=\frac{\pi}{2}$, with $\delta$ either $0$ or $\pi$,  
is a 1-dimension line in physical $ V$ space,  given by \eqref{eq:V20}
with $0\le \theta_+\le \frac{\pi}{2}$, and there are only two physically distinct fixed points where the
$\beta$-functions vanish, \eqref{eq:theta-dot} and \eqref{eq:delta-dot} , 
$\theta_+=0$ and $\theta_+=\frac{\pi}{2}$.\footnote{This line is a RG invariant space in 
$\bigl(U(1)\times U(1)\bigr)\backslash SU(3)/\bigl(U(1)\times U(1)\bigr)$, it is a  double coset  
$U(1)\backslash SU(2)/U(1)$.}
This gives the two distinct fixed points when $\theta_2=\frac{\pi}{2}$ as
\[
 V^*=
  \begin{pmatrix}
0&0&1\\ 0&1&0\\ 1 &0&0
  \end{pmatrix}
\quad\mbox{and} \quad 
\begin{pmatrix}
  0&0&1\\ 1   &0&0\\ 0 &1&0
\end{pmatrix},
\] 
 where all entries have been rendered positive by phase transformations of the fermions.

Although there  are a total of ten parameters in the Yukawa sector,
it makes physical sense to focus on the the fixed points of the mixing parameters alone.
Label the six Yukawa couplings by $y_\alpha=(y_a,y_a')$, $\alpha=1,\ldots,6$, and the four mixing parameters by $\vartheta_\mu= (\theta_1,\theta_2,\theta_3,\delta)$, $\mu=1,\ldots, 4$.
   A fixed point of the full Yukawa sector  requires $\beta_\alpha=\frac{d y_\alpha}{d t}=0$ and 
$\beta_\mu=\frac{d \vartheta_\mu}{d t}=0$.   But the six fixed points of $\beta_\mu$, determined by
\eqref{eq:theta-dot} and \eqref{eq:delta-dot}, require all the $P$s and $Q$s in appendix \ref{app:PQ} to vanish, regardless of the prefactors, so if $\vartheta_\mu$ are chosen so that $\beta_\mu(\vartheta)$ =0 for some $y_\alpha$, then  $\beta_\mu(\vartheta)$  =0 for all $y_\alpha$, independently of the values of the $\beta_\alpha$.  
It is shown in appendix \ref{app:lambda} that the $10\times 10$ matrix of anomalous dimensions is block diagonal when  
\eqref{eq:theta-dot} and \eqref{eq:delta-dot} vanish, so it makes physical sense to view  the above six points where $\beta_\mu=0$  to be fixed points of the mixing matrix flow, independently of the Yukawa coupling $\beta$-functions $\beta_\alpha$.

\begin{figure}
{\centering 
\includegraphics[scale =0.5]{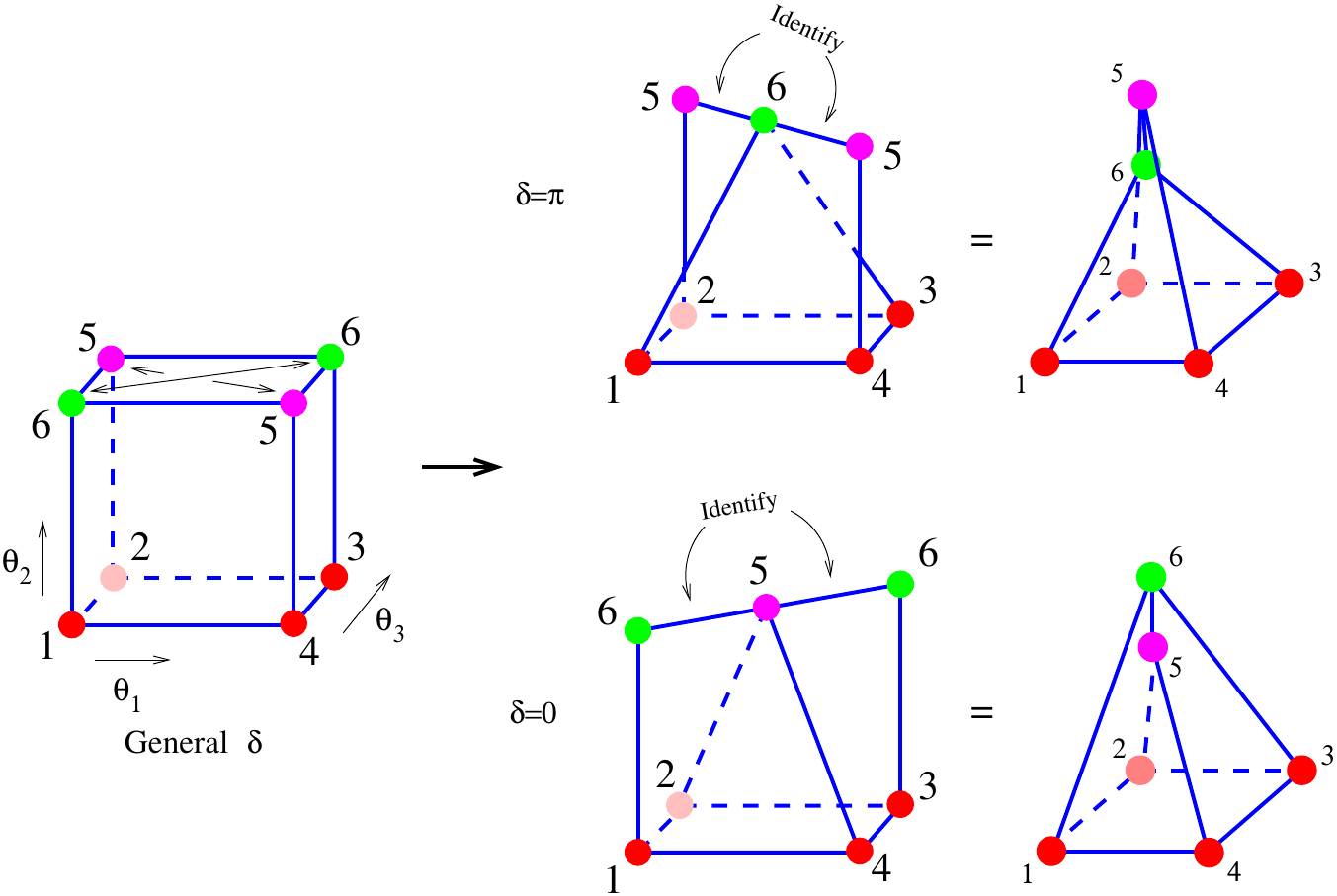} 

}
\caption{When $\delta=0$ or $\pi$ the  $\theta_{2}=\frac{\pi}{2}$ face of the cube degenerates to a line segment which is topologically equivalent to the line on the right of  figure \ref{fig:SU2}.}
 \label{fig:SU3}
\end{figure}

In summary, there are six fixed points of the 1-loop RG flow of the mixing matrix,
and $V$ at these points is (up to phase transformations)
{\scriptsize
\begin{alignat}{4}
    V_1^*=&\begin{pmatrix} 1&0&0\\ 0&1&0\\ 0 &0&1 \end{pmatrix},&
             \   & (\theta_{1}^*,\theta_{2}^*,\theta_{3}^*)=(0,0,0);&     \ 
 V^*_2=&\begin{pmatrix} 0&1&0\\ 1&0&0\\ 0 &0&1 \end{pmatrix},&
       \   &(\theta_{1}^*,\theta_{2}^*,\theta_{3}^*)=\Bigl(0,0,\frac{\pi}{2}\Bigr);\nonumber\\
      V_3^* =  &\begin{pmatrix} 0&1&0\\ 0&0&1\\ 1&0&0 \end{pmatrix},&
  \    &(\theta_{1}^*,\theta_{2}^*,\theta_{3}^*)= \Bigl(\frac{\pi}{2},0,\frac{\pi}{2}\Bigr);&  \ 
 V^*_4=&\begin{pmatrix} 1&0&0\\ 0&0&1\\ 0 &1&0  \end{pmatrix},&
     \  &(\theta_{1}^*,\theta_{2}^*,\theta_{3}^*)=\Bigl(\frac{\pi}{2},0,0\Bigr);\nonumber\\
 V_5^*=&  \begin{pmatrix} 0&0&1\\ 1  &0&0\\ 0 &1&0 \end{pmatrix},&
       \      &  (\theta_{1}^*,\theta_{2}^*,\theta_{3}^*)=
          \begin{cases}
            \Bigl(0,\frac{\pi}{2},\frac{\pi}{2}\Bigr),\\[0.5em]
            \Bigl(\frac{\pi}{2},\frac{\pi}{2},0\Bigr) ;
          \end{cases}&         \  
   V^*_6 =& \begin{pmatrix} 0&0&1\\ 0&1&0\\ 1 &0&0 \end{pmatrix},&
                  \  & (\theta_{1}^*,\theta_{2}^*,\theta_{3}^*)
                  =\begin{cases}
                   \Bigl(0,\frac{\pi}{2},0\Bigr),\\[0.5em]
               \Bigl(\frac{\pi}{2},\frac{\pi}{2},\frac{\pi}{2}\Bigr).
               \end{cases}\label{eq:S_3}
\end{alignat}
}
These are labeled  1 to 6 in the above order:
with this labeling of the fixed points the situation can be visualised as  shown in figure \ref{fig:SU3}.

The six matrices $V_1^*-V_6^*$ actually form a group, they furnish a unitary representation of the symmetric group acting on three objects.  The appearance of this group is
related to the fact that $S_3$  is the Weyl group of $SU(3)$ and these six points are also fixed points of the left action of the Cartan torus of $SU(3)$, $T\approx U(1)\times U(1)$, on the flag manifold $F_3$ .\footnote{The fact that there are 6 fixed points is related to fixed point theorems: the Euler characteristic of $F_3$ is 6.
I am grateful  to Charles Nash for bringing my attention to fixed point theorems and the significance of the Weyl group of $SU(3)$ in this context.}

 Note that the Jarlskog invariant \cite{Jarlskog}
\begin{align*}{\cal J}=\frac{i}{4}\det[Z',Z]&=(y^2_3-y^2_2)(y^2_2-y^2_1)(y^2_1-y^2_3)(y^{\prime 2}_3-y^{\prime 2}_2)(y^{\prime 2}_2-y^{\prime 2}_1)(y^{\prime 2}_1-y^{\prime 2}_3)J,\\
  \mbox{with} \quad   J &=  \frac{1}{8}\sin 2\theta_{1} \sin 2 \theta_{2} \sin 2\theta_{3} \cos\theta_{2}\sin(\phi_{1}+\phi_{2}+\phi_{3})\\
        &= \frac{1}{8}\sin 2\theta_{1} \sin 2 \theta_{2} \sin 2\theta_{3} \cos\theta_{2}\sin\delta,
      \end{align*}
vanishes at all of the fixed points.
 
\section{Operator mixing \label{sec:AD}}

In any quantum field theory, with couplings $x_A$  and $\beta$-functions 
$\beta_A = \frac{ d x_A}{d t}$ with fixed points $x_A^*$, operator mixing under the RG flow is controlled by the  matrix $\Gamma$ with entries
\[\Gamma_A{}^B =  \bpm \frac{\partial \beta_A}{\partial x_B}\epm.\] 
The nature of each fixed point,
whether attractive, repulsive or unstable,  is determined by the signs of the eigenvalues of  $\Gamma$ at the fixed point.  Although the components of $\Gamma_*$ depend on the parameterisation $x_A$ its eigenvalues do not, they are invariant under a reparameterisation, $x_A \rightarrow x'_A(x)$, and hence are not scheme dependent.

It is shown in appendix \ref{app:lambda} that, at the six fixed points,  $\vartheta^*$,  of the fermion mixing matrix, the $10\times 10$ matrix $\Gamma$ is block diagonal at 1-loop, 
\beq 
\Gamma_A{}^B =\bpm \frac{\partial \beta_\alpha}{\partial y_\beta} &  0 \\  
      0 &  \left.\frac{\partial \beta_\mu}{\partial \vartheta_\nu} \right|_*\epm \label{eq:block}\eeq
for any values of $y_\alpha$, not just at $y_\alpha^*$.
It therefore  makes sense to focus on\footnote{$\gamma_*$ is the matrix of anomalous dimensions, 
but since angles are necessarily dimensionless, this may not be an appropriate name in this context.}
     \[ \gamma_* =
       \left.\bpm
       \frac{\partial \beta_{1}}{\partial\theta_{1}}&
       \frac{\partial \beta_{1}}{\partial\theta_{2}}&
       \frac{\partial \beta_{1}}{\partial\theta_{3}}&
           \frac{\partial \beta_{1}}{\partial\delta}\\[0.2em]
       \frac{\partial \beta_{2}}{\partial\theta_{1}}&
       \frac{\partial \beta_{2}}{\partial\theta_{2}}&

       \frac{\partial \beta_{2}}{\partial\theta_{3}}&
       \frac{\partial \beta_{2}}{\partial\delta}\\[0.2em]
       \frac{\partial \beta_{3}}{\partial\theta_{1}}&
       \frac{\partial \beta_{3}}{\partial\theta_{2}}&
       \frac{\partial \beta_{3}}{\partial\theta_{3}}&
           \frac{\partial \beta_{3}}{\partial\delta}\\[0.2em]
       \frac{\partial \beta_\delta }{\partial\theta_{1}}&
       \frac{\partial \beta_\delta }{\partial\theta_{2}}&
       \frac{\partial \beta_\delta }{\partial\theta_{3}}&
           \frac{\partial \beta_\delta }{\partial\delta}
           \epm\right|_* \]
separately.

A subtlety is that for $\theta_{2}=\frac{\pi}{2}$, and $\delta=0$ or $\pi$, $\frac{\partial \beta_\delta }{\partial \delta} $ is  indeterminate, because the $(\theta_{1},\theta_{2},\theta_{3},\delta)$ coordinates are singular there: better is to change coordinates to $(\theta_{1},\theta_{2},\theta_{3},J)$ to use
  \beq\gamma_* =
      \left. \bpm
       \frac{\partial \beta_{1}}{\partial\theta_{1}}&
       \frac{\partial \beta_{1}}{\partial\theta_{2}}&
       \frac{\partial \beta_{1}}{\partial\theta_{3}}&
           \frac{\partial \beta_{1}}{\partial J }\\[0.2em]
       \frac{\partial \beta_{2}}{\partial\theta_{1}}&
       \frac{\partial \beta_{2}}{\partial\theta_{2}}&
       \frac{\partial \beta_{2}}{\partial\theta_{3}}&
       \frac{\partial \beta_{2}}{\partial J }\\[0.2em]
       \frac{\partial \beta_{3}}{\partial\theta_{1}}&
       \frac{\partial \beta_{3}}{\partial\theta_{2}}&
       \frac{\partial \beta_{3}}{\partial\theta_{3}}&
           \frac{\partial \beta_{3}}{\partial J }\\[0.2em]
       \frac{\partial \beta_J }{\partial\theta_{1}}&
       \frac{\partial \beta_J }{\partial\theta_{2}}&
       \frac{\partial \beta_J }{\partial\theta_{3}}&
           \frac{\partial \beta_J }{\partial J}
           \epm\right|_*, \label{eq:gamma*}\eeq
where $\beta_J = \frac{d J}{d t}$, and  then $\left.\frac{\partial \beta_J }{\partial J}\right|_*$ is well defined for $\theta_{2}=\frac{\pi}{2}$ and $\delta=0$ or $\pi$. 
 The eigenvalues of $\gamma_*$
at the six fixed points (at 1-loop) are given in appendix  \ref{app:lambda}.

In the following subsection we shall examine $\gamma_*$ in more detail  for the quark sector of the Standard Model.

\subsection{The CKM matrix}
In the quark sector the Standard Model hierarchy, with $y_u^2 \ll y_c^2 \ll y_t^2$ and
\hbox{$ y_d^2 \ll y_s^2  \ll y_b^2 $},  it is a good approximation to  set \hbox{$z_i=z'_i=1$}.
In this limit \eqref{eq:theta-dot} and \eqref{eq:delta-dot} reduce to 
\begin{align}
  \beta_{1} &=   \frac{3 \sin2\theta_{1}  }{64 \pi^2}
                     \bigl(y_t^2+  y_b^2\cos^2\theta_{2} \bigr),  \label{eq:dottheta23}\\[1em]
           \beta_{2}  &=
                                \frac{3   \sin 2\theta_{2}  }{64 \pi^2 } \bigl(y_t^2 \cos^2 \theta_{1}+ y_b^2  \bigr),
                               \label{eq:dottheta31}\\[1em]
        \beta_{3} & =
    \frac{3 y_t^2 }{64 \pi^2 }
     \Bigl[ \sin 2 \theta_{3}(\sin^2\theta_{1}   - \cos^2 \theta_{1}  \sin^2\theta_{2}   )
                           - 2\cos \delta   \cos^2\theta_{3}\sin 2 \theta_{1}\sin \theta_{2} \Bigr],\\
        \beta_\delta   &= \frac{3 y_t^2}{64 \pi^2 }\sin \delta   \cot  \theta_{3} 
                 \sin 2 \theta_{1} \sin \theta_{2}.\label{eq:dotdelta}
\end{align}

Assuming the above hierarchy persists under RG flow of the Yukawa couplings, the signs of the eigenvalues of the operator mixing matrix at the six fixed points can be found from the expressions in appendix \ref{app:lambda} with  $y_t^2  \gg y_c^2  \gg y_u^2 $, $y_b^2\gg y_s^2 \gg y_d^2$ and  $z_i=z_i'=1$ (so $\zeta_{ij}=\zeta'_{ij}=\frac{1}{2}$ in appendix \ref{app:lambda}).  Defining  $\widehat \lambda_i = \frac{32 \pi^2}{3} \lambda_i$ the eigenvalues in the UV-direction, to leading order, are:
\begin{center}
  \begin{tabular}{ | c |  c   c   c  c   | }
    \hline
   &&&& \\[-1em]
  Fixed Point
    & $\widehat\lambda_{1}$
    & $\widehat \lambda_{2}$
    &  $\widehat \lambda_{3}$
    &  $\widehat \lambda_J$ \\[0.2em]
    \hline
      &&&& \\[-1em] 
   1                                           & $y_t^2+y_b^2 $ &  $y_t^2+y_b^2$ & $0$ &  $2 (y_t^2+y_b^2)$  \\ [0.1em]
 2                 & $y_t^2+y_b^2$ &  $y_t^2+y_b^2$ & $0$ &  $2 (y_t^2+y_b^2)$   \\[0.1em]
  3 &  $-(y_t^2+y_b^2) $ &   $y_b^2$ &  $-y_t^2$ &    $-2 y_t^2$\\[0.2em]
 4            &  $-(y_t^2+y_b^2 )$ & $y_b^2$ &  $y_t^2$ &   $0$  \\[0.1em]
    \hline
  \end{tabular}
\end{center}
for the fixed points 1-4.
For the fixed points 5 and 6 it is better to rotate the coordinates in the $\theta_1-\theta_3$ plane to $\theta_+ = \theta_3 + \theta_1$ and $\theta_-=\theta_3 - \theta_1$ (see appendix \ref{app:lambda}) and $\lambda_\pm = \left.\frac{\partial \dot\theta_\pm}{\partial \theta_\pm}\right|_*$ are eigenvalues.  For the fixed point 5, the directions along the edges
25 and 45 in figure 2 are eigendirections near the point 5;
and for the fixed point 6, the directions along the edges
16 and 36  are eigendirections near the point 6: the corresponding eigenvalues are labelled
$\lambda_{25}$, $\lambda_{45}$, $\lambda_{16}$ and $\lambda_{36}$ in the tables below:
\begin{center}
  \begin{tabular}{ | c  |c   c   c  c   | }
    \hline
   &&& &\\[-1em]
  Fixed Point
    & $\widehat \lambda_+$
    & $\widehat \lambda_{25}$
                                    &  $\widehat{\lambda}_{45}$
                                          &  $\widehat \lambda_J$ \\[0.2em]
    \hline
    &&&& \\[-1em]  
5& $ y_t^2$ &   $-y_b^2$ &   $-(y_t^2+y_b^2)$  &    $-2 y_b^2$\\[0.2em]
    \hline
  \end{tabular}
\end{center}
\begin{center}
  \begin{tabular}{ | c  |c   c   c  c   | }
    \hline
   &&& &\\[-1em]
  Fixed Point
    & $\widehat \lambda_-$
    & $\widehat \lambda_{16}$
                                    &  $\widehat{\lambda}_{36}$
                                          &  $\widehat \lambda_J$ \\[0.2em]
    \hline
    &&&& \\[-1em]  
  6 &  $-y_t^2 $ &   $-(y_t^2+y_b^2)$ &  $-y_b^2$ &    $-2(y_t^2+y_b^2)$\\[0.2em]
    \hline
  \end{tabular}
\end{center}
for the fixed points 5 and 6.

One could fill in the zeros by relaxing the conditions $z_i=z_i'=1$.
For example letting $z_1$ and $z_1'$ deviate slightly from unity, keeping $z_2=z_3 =z'_2=z'_3=1$, gives  $\lambda_3= \frac{3 }{32\pi^2}(y_c^2 + y_s^2) $ at the fixed point 1
and $\lambda_3= -\frac{3 }{32 \pi^2}(y_c^2 +y_s^2) $ at the fixed point 2:  in particular this choice makes fixed point 1  fully repulsive  in the UV direction,  but higher loops could modify  this conclusion.
Thus, in this limit, fixed point 6 is fully attractive in the UV  while fixed point 5 is fully attractive in the IR.

\section{Fixed points at all orders \label{sec:allorders}}

It is not a coincidence that the 6 fixed points of the 1-loop RG
equations for the mixing angles and phase coincide with the 6 elements of the Weyl group of $SU(3)$ that are fixed points of the left action of the Cartan torus $T$ on the flag manifold $F_3$: indeed  this must be true to all orders in perturbation theory, and even non-perturbatively, if it is  assumed that the action of  $T$ on  $F_3$ commutes with the RG flow on  $F_3$.

Let $\Theta=(\theta_1,\theta_2,\theta_3,\phi_1,\phi_2,\phi_3)$ be a point in $F_3$ in an arbitrary gauge (in the sense of the energy dependence of the fermion phases), and let $\varphi(\Theta)$ be an infinitesimal RG transformation of $\Theta$.     Let $\Theta^*$ be a fixed point of the action of $T$ on $F_3$,  so ${\mathbf t} (\Theta^*)=\Theta^*$  for some ${\mathbf t} \in T$.   If $\Theta^*$ were 
not a RG fixed point,  it could  be moved to an infinitesimally
close point $\varphi(\Theta^*) \ne \Theta^*$ with a RG transformation, $\varphi$.  Demanding that
${\mathbf t}  \circ \varphi(\Theta^*)
= \varphi \circ {\mathbf t} (\Theta^*) 
= \varphi(\Theta^*)$ then implies that 
$\varphi(\Theta^*)$  is also left invariant by ${\mathbf t} $. But the 6 fixed points
of $T$ are isolated, so there are no other fixed points of $T$ infinitesimally close to $\Theta^*$. Hence
$\varphi(\Theta^*) = \Theta^*$
and $\Theta^*$ must be a RG fixed point.\footnote{Again I thank Charles Nash for pointing out the significance of the fact that these actions commute for the fixed points.
The version of the argument presented here was used in the context of modular symmetry and the quantum Hall effect in \cite{QHE}.}
This must even be true non-perturbatively; all that has been assumed is that  $T$ commutes
with the RG flow on $F_3$.

This argument does not preclude the possibility that there are fixed points of the RG that are not fixed points of the left action of $T$.  But, if the Weyl group also commutes with the RG flow, then  any such fixed points must come in sets of six.  The argument is also only valid for Dirac fermions: for Majorana fermions there are more parameters and bare masses that are not considered in the present work (see \cite{DIS} and \cite{RRS} for partial 2-loop corrections with Majorana neutrinos).\footnote{I thank the referee for pointing out these references.}

\section{Discussion \label{sec:conclusions}}

It has been shown that 1-loop running of the mixing angles and the CP-violating phase
for the Standard Model with three generations gives rise to six RG fixed points.
These six points are related  to the Weyl group of $SU(3)$, whose elements correspond to fixed points of the left action of the Cartan torus on $F_3$.  The same analysis applies to both the quark sector and the leptonic sector with three right-handed, gauge singlet,  Weyl (but not Majorana) neutrinos. 

Only massless RG running has been considered here, so the present analysis only applies to energies significantly above the Higgs and $t$-quark masses, a few hundred GeV and higher.  In the IR direction one would need to include the effect of masses and take into account the fact that particles drop out of the running as the energy falls below their mass threshold.  It is not the case that $\theta_1, \theta_2$ and $\theta_3$ are all driven to zero in the IR in the Standard Model hierarchy, rather they will freeze at their observed values for energies below a few hundred GeV.
In any case the analysis presented here is not directly relevant to the physics of the Standard Model.   The observed magnitudes of the Yukawa couplings are so small 
that the 1-loop running of the CKM parameters in the UV direction is so slow that there are no significant physical effects even up to the largest conceivable energies.

Nevertheless the analysis could be useful for more formal aspects of the theory.
For example the idea of gradient flow
\cite{Wallace+Zia} requires introducing a metric on the space of couplings and 
a natural metric on the double $SU(3)$ coset
\hbox{$T\backslash F_3$} could be obtained from the restriction of the $SU(3)$-invariant metric on the flag manifold,
this will be direction of future investigations.

The ideas presented here could also be relevant to models beyond the Standard Model:
dark matter could arise from  a gauge theory, similar to the Standard Model,
with a portal to the Standard Model Higgs but with no coupling to the photon.  If there were $N_g$ dark generations of chiral fermions with Yukawa coupling matrices, the  mixing angles and CP-violating phases would be parameterised
by $U(1)^{N_g-1}\backslash SU(N_g) /U(1)^{N_g-1}$ which has dimension $(N_g-1)^2$,
consisting of $\frac{1}{2} N_g (N_g-1)$ angles and 
\hbox{$\frac{1}{2}(N_g-1)(N_g-2)$} CP-violating phases.
The left action of $U(1)^{N_g-1}$  has $N_g!$ fixed points, corresponding to the Weyl group of  $SU(N_g)$ (which is the symmetric group acting on $N_g$ objects, $S_{N_g}$) and these will be fixed points of the RG.
If the relevant Yukawa couplings were large enough the system could be driven in the infra-red toward significant CP-violation before the running cuts off due to masses,
satisfying one of the conditions for the observed CP-violation in our Universe.

The present work has not considered Majorana fermions, which would introduce more parameters and masses.  Such an  analysis would be more involved but, at least in principle,  could be tackled using similar techniques: this is left for future work.

It is a pleasure to thank Charles Nash and Denjoe O'Connor for useful conversations
on invariants of multi-matrix systems and fixed points.

The manipulations necessary to derive the polynomials $P$ and $Q$ and the matrix of anomalous dimensions were all performed using the symbolic manipulation software Mathematica\textsuperscript{TM}.

\newpage
\appendix

\section{Feynman diagrams \label{app:Feynman}}

The 1-loop Feynman diagrams that contribute to the running of  the Yukawa couplings
are shown below:

\bigskip

  \begin{tikzpicture}
    \begin{feynman}
        \vertex at (0,0) (a) ;   
        \vertex at (1.5,0) (b);
                  \diagram*{
            (a) -- [gluon, edge label=\({\scriptstyle SU(3)}\)] (b),
                       };
    \vertex at (3,0) (c) ;   
        \vertex at (4.5,0) (d);
                  \diagram*{
    (c) -- [boson, color=red,very thick,edge label=\(\textcolor{black}{\scriptstyle SU(2)}\)] (d),
      };
 \vertex at (6,0) (e) ;   
        \vertex at (7.5,0) (f);
                  \diagram*{
            (e) -- [photon, edge label=\({\scriptstyle U(1)}\)] (f),
                       };
                \end{feynman}
              \end{tikzpicture}
   
            \bigskip

           \begin{tikzpicture}
    \begin{feynman}
        \vertex at (0, 1) (fR)  {\(q_R\)};   
        \vertex at (0,-1) (fL)  {\(q_L\)};
        \vertex at (1,0) (v);
        \vertex at (2,0) (h);
\vertex at (0.9,0.1)(b);
\vertex at (0.5,0.5)(a);
                  \diagram*{
            (fL) -- [fermion] (v) -- [plain] (b)  -- [fermion](a) ,
(a) -- [plain](fR),
 (b) -- [ gluon, half right] (a),
            (v) -- [scalar] (h),
                       };
                     \end{feynman}
                   \end{tikzpicture}
                 \begin{tikzpicture}
    \begin{feynman}
        \vertex at (0, 1) (fR)  {\(q_R\)};   
        \vertex at (0,-1) (fL)  {\(q_L\)};
        \vertex at (1,0) (v);
        \vertex at (2,0) (h);
\vertex at (0.9,-0.1)(b);
\vertex at (0.5,-0.5)(a);
                  \diagram*{
            (fL) --  [plain] (b)  -- [anti fermion](a)  --[plain] (v) ,
(v) -- [fermion](fR),
 (a) -- [ gluon, half right] (b),
            (v) -- [scalar] (h),
                       };
                     \end{feynman}
                   \end{tikzpicture}
  \begin{tikzpicture}
    \begin{feynman}
        \vertex at (0, 1) (fR)  {\(q_R\)};   
        \vertex at (0,-1) (fL)  {\(q_L\)};
        \vertex at (1,0) (v);
        \vertex at (2,0) (h);
\vertex at (0.9,-0.1)(b);
\vertex at (0.5,-0.5)(a);
\vertex at (0.9,0.1)(d);
\vertex at (0.5,0.5)(c);
                  \diagram*{
            (fL) --  [plain] (b)  -- [anti fermion](a)  --[plain] (v) ,
            (v) -- [plain] (d) --[fermion](c) --[plain](fR),
            (c) -- [gluon](a),
            (v) -- [scalar] (h),
                       };
                     \end{feynman}
                   \end{tikzpicture}
                      \begin{tikzpicture}
    \begin{feynman}
        \vertex at (0, 1) (fR)  {\(f_R\)};   
        \vertex at (0,-1) (fL)  {\(f_L\)};
  \vertex at (1,0) (v);
  \vertex at (2.2,0) (h);
\vertex at (1.5,0)(a);
\vertex at (1.5,0.6)(b);
                  \diagram*{
            (fL) -- [fermion] (v) ,
            (v)  --[fermion] (fR),
            (v) -- [scalar](a) -- [scalar](h),
%
               (a) -- [ gluon, quarter right] (b),
         (b) -- [ gluon, quarter right] (a),
};
                     \end{feynman}
                   \end{tikzpicture}
                   
 \begin{tikzpicture}
   \begin{feynman}
        \vertex at (0, 1) (fR)  {\(f_R\)};   
        \vertex at (0,-1) (fL)  {\(f_L\)};
        \vertex at (1,0) (v);
        \vertex at (2,0) (h);
\vertex at (1.5,0)(c);
\vertex at (0.9,-0.1)(b);
\vertex at (0.5,-0.5)(a);
                  \diagram*{
            (fL) --  [plain] (b)  -- [anti fermion](a)  --[plain] (v) ,
(v) -- [fermion](fR),
 (a) -- [ boson, color=red,very thick,half right] (c),
            (v) -- [scalar](c) -- [scalar] (h),
                       };
                     \end{feynman}
                   \end{tikzpicture}
                        \begin{tikzpicture}
     \begin{feynman}
        \vertex at (0, 1) (fR)  {\(f_R\)};   
        \vertex at (0,-1) (fL)  {\(f_L\)};
        \vertex at (1,0) (v);
        \vertex at (2,0) (h);
\vertex at (0.9,-0.1)(b);
\vertex at (0.5,-0.5)(a);
                  \diagram*{
            (fL) --  [plain] (b)  -- [anti fermion](a)  --[plain] (v) ,
(v) -- [fermion](fR),
 (a) -- [boson, color=red, very thick, half right] (b),
            (v) -- [scalar] (h),
                       };
                     \end{feynman}
                   \end{tikzpicture}
                 \begin{tikzpicture}
         \begin{feynman}
        \vertex at (0, 1) (f2)  {\(f_R\)};
        \vertex at (0,-1) (f1)  {\(f_L\)};
        \vertex at (1,0) (a);
        \vertex at (2.5,0) (h);
\vertex at (1.3,0)(b);
\vertex at (2.2,0)(c);
                  \diagram*{
            (f1) -- [fermion] (a)  -- [fermion](f2),
            (a) -- [scalar] (b) -- [scalar] (c) -- [scalar] (h),
       (b) -- [boson, color=red, very thick, half left] (c), 
                  };
    \end{feynman}
  \end{tikzpicture}
    \begin{tikzpicture}
    \begin{feynman}
        \vertex at (0, 1) (fR)  {\(f_R\)};   
        \vertex at (0,-1) (fL)  {\(f_L\)};
  \vertex at (1,0) (v);
  \vertex at (2,0) (h);
\vertex at (1.5,0)(a);
\vertex at (1.5,0.7)(b);
                  \diagram*{
            (fL) -- [fermion] (v) ,
            (v)  --[fermion] (fR),
            (v) -- [scalar](a) -- [scalar](h),
            (a) -- [ boson, color=red,very thick,half right] (b),
             (a) -- [ boson, color=red,very thick,half left] (b),
         };
                     \end{feynman}
                   \end{tikzpicture}
                   
                 \begin{tikzpicture}
    \begin{feynman}
        \vertex at (0, 1) (fR)  {\(f_R\)};   
        \vertex at (0,-1) (fL)  {\(f_L\)};
        \vertex at (1,0) (v);
        \vertex at (2,0) (h);
\vertex at (1.5,0)(c);
\vertex at (0.9,-0.1)(b);
\vertex at (0.5,-0.5)(a);
                  \diagram*{
            (fL) --  [plain] (b)  -- [anti fermion](a)  --[plain] (v) ,
(v) -- [fermion](fR),
 (a) -- [photon, half right] (c),
            (v) -- [scalar](c) -- [scalar] (h),
                       };
                     \end{feynman}
                   \end{tikzpicture}
                   \begin{tikzpicture}
    \begin{feynman}
        \vertex at (0, 1) (fR)  {\(f_R\)};   
        \vertex at (0,-1) (fL)  {\(f_L\)};
        \vertex at (1,0) (v);
        \vertex at (2,0) (h);
\vertex at (1.5,0)(c);
\vertex at (0.9,0.1)(b);
\vertex at (0.5,0.5)(a);
                  \diagram*{
            (fL) -- [fermion] (v) ,
(v) -- [fermion](fR),
 (a) -- [photon, half left] (c),
            (v) -- [scalar](c) -- [scalar] (h),
                       };
                     \end{feynman}
                   \end{tikzpicture}
                 \begin{tikzpicture}
    \begin{feynman}
        \vertex at (0, 1) (fR)  {\(f_R\)};   
        \vertex at (0,-1) (fL)  {\(f_L\)};
        \vertex at (1,0) (v);
        \vertex at (2,0) (h);
\vertex at (0.9,-0.1)(b);
\vertex at (0.5,-0.5)(a);
\vertex at (0.9,0.1)(d);
\vertex at (0.5,0.5)(c);
                  \diagram*{
            (fL) --  [plain] (b)  -- [anti fermion](a)  --[plain] (v) ,
            (v) -- [plain] (d) --[fermion](c) --[plain](fR),
            (c) -- [photon](a),
            (v) -- [scalar] (h),
                       };
                     \end{feynman}
                   \end{tikzpicture}  
                 \begin{tikzpicture}
    \begin{feynman}
        \vertex at (0, 1) (fR)  {\(f_R\)};   
        \vertex at (0,-1) (fL)  {\(f_L\)};
        \vertex at (1,0) (v);
        \vertex at (2,0) (h);
\vertex at (0.9,-0.1)(b);
\vertex at (0.5,-0.5)(a);
                  \diagram*{
            (fL) --  [plain] (b)  -- [anti fermion](a)  --[plain] (v) ,
            (v) -- [fermion](fR),
 (a) -- [photon, half right] (b),
            (v) -- [scalar] (h),
                       };
                     \end{feynman}
                   \end{tikzpicture}
                   \begin{tikzpicture}
                     \begin{feynman}
        \vertex at (0, 1) (fR)  {\(f_R\)};   
        \vertex at (0,-1) (fL)  {\(f_L\)};
        \vertex at (1,0) (v);
        \vertex at (2,0) (h);
\vertex at (0.9,0.1)(b);
\vertex at (0.5,0.5)(a);
                  \diagram*{
            (fL) -- [fermion] (v) -- [plain] (b)  -- [fermion](a) ,
(a) -- [plain](fR),
 (a) -- [ photon, half left] (b),
            (v) -- [scalar] (h),
                       };
                     \end{feynman}
                   \end{tikzpicture}
                   
\begin{tikzpicture}
         \begin{feynman}
        \vertex at (0, 1) (f2)  {\(f_R\)};
        \vertex at (0,-1) (f1)  {\(f_L\)};
        \vertex at (1,0) (a);
        \vertex at (2.5,0) (h);
\vertex at (1.3,0)(b);
\vertex at (2.2,0)(c);
                  \diagram*{
            (f1) -- [fermion] (a)  -- [fermion](f2),
            (a) -- [scalar] (b) -- [scalar] (c) -- [scalar] (h),
       (b) -- [photon, half left] (c) ,
                  };
    \end{feynman}
  \end{tikzpicture}
\begin{tikzpicture}
         \begin{feynman}
        \vertex at (0, 1) (f2)  {\(f_R\)};
        \vertex at (0,-1) (f1)  {\(f_L\)};
        \vertex at (1,0) (a);
        \vertex at (2.5,0) (h);
\vertex at (1.4,0)(b);
\vertex at (2.0,0)(c);
                  \diagram*{
            (f1) -- [fermion] (a)  -- [fermion](f2),
            (a) -- [scalar] (b) ,
            (c) -- [scalar] (h),
            (b) -- [fermion, half left] (c) ,
            (c) -- [fermion, half left] (b) ,
                  };
    \end{feynman}
  \end{tikzpicture}
      \begin{tikzpicture}
    \begin{feynman}
        \vertex at (0, 1) (fR)  {\(f_R\)};   
        \vertex at (0,-1) (fL)  {\(f_L\)};
        \vertex at (1,0) (v);
        \vertex at (2,0) (h);
\vertex at (0.9,-0.1)(b);
\vertex at (0.5,-0.5)(a);
\vertex at (0.9,0.1)(d);
\vertex at (0.5,0.5)(c);
                  \diagram*{
            (fL) --  [plain] (b)  -- [anti fermion](a)  --[plain] (v) ,
            (v) -- [plain] (d) --[fermion](c) --[plain](fR),
            (c) -- [scalar](a),
            (v) -- [scalar] (h),
                       };
                     \end{feynman}
                   \end{tikzpicture}  
   \begin{tikzpicture}
    \begin{feynman}
        \vertex at (0, 1) (fR)  {\(f_R\)};   
        \vertex at (0,-1) (fL)  {\(f_L\)};
  \vertex at (1,0) (v);
  \vertex at (2,0) (h);
\vertex at (1.5,0)(a);
\vertex at (1.5,0.7)(b);
                  \diagram*{
            (fL) -- [fermion] (v) ,
            (v)  --[fermion] (fR),
            (v) -- [scalar](a) -- [scalar](h),
           (a) -- [scalar, half left ] (b),
          (a) -- [scalar, half right ] (b),
         };
                     \end{feynman}
                   \end{tikzpicture}

                 \newpage

\section{The polynomials $\bm P$ and $\bm Q$ \label{app:PQ}}

The polynomial $P$ and $Q$ in equations  \eqref{eq:theta-dot} and \eqref{eq:delta-dot} are
 %
            \begin{alignat*}{3}
      &           P^{1  1}_{1} && =&& \sin 2\theta_{1}\cos^2 \theta_{3}
                              - \cos \delta  \sin ^2\theta_{1} \sin \theta_{2} \sin 2\theta_{3},\\
  &   P^{21}_{1} &&=&&  \sin 2\theta_{1}\sin^2\theta_{3}
                              +  \cos \delta  \sin ^2\theta_{1} \sin \theta_{2} \sin2 \theta_{3},\\
     &      P^{1  2}_{1}      && =  &&\cos \delta   \sin 2 \theta_{3}\sin \theta_{2},\\
                     &      P^{22}_{1}      && = - &&\cos \delta   \sin 2 \theta_{3}\sin \theta_{2},\\
                         &       Q^{11}_{1} && =&& \sin 2\theta_{1}
                                      (\cos^2 \theta_{3}-\sin^2\theta_{2}\sin^2\theta_{3} )
                                      +  \cos \delta \cos 2\theta_{1}\sin \theta_{2}  \sin 2 \theta_{3},\\
   &             Q^{21}_{1}   && =&& \sin 2\theta_{1} \sin^2 \theta_{2} \sin ^2\theta_{3}+\cos \delta  \sin^2 \theta_{1} \sin \theta_{2} \sin 2 \theta_{3},\\
     &   Q^{31}_{1}  && = - &&\sin 2\theta_{1} \sin^2 \theta_{2} \sin ^2\theta_{3}
                                 +\cos \delta  \cos^2 \theta_{1} \sin \theta_{2}  \sin 2 \theta_{3},\\
          &      Q^{12}_{1}   && =&& \sin 2\theta_{1}
                                      (\sin ^2\theta_{3}- \sin^2\theta_{2}\cos^2 \theta_{3})
                                      - \cos \delta  \cos 2\theta_{1} \sin \theta_{2} \sin 2 \theta_{3} ,\\
       &   Q^{22}_{1}   && =&& \sin 2 \theta_{1} \sin ^2\theta_{2} \cos ^2\theta_{3}
                                      - \cos \delta  \sin^2 \theta_{1}  \sin \theta_{2} \sin 2 \theta_{3}, \\
 &     Q^{32}_{1}    && = -&&\sin 2\theta_{1} \sin^2 \theta_{2} \cos ^2\theta_{3}
                              - \cos \delta  \cos^2 \theta_{1}    \sin \theta_{2} \sin 2 \theta_{3} .\\[1em]
     %
   &        P^{1  1}_{2}&&=  - &&\sin 2\theta_{2} \sin ^2\theta_{3}\sin ^2\theta_{1} 
           + \frac 1 2 \cos \delta   \cos \theta_{2} \sin 2 \theta_{3} \sin 2\theta_{1},\\
     &      P^{21}_{2}&&=-   &&\sin 2\theta_{2}   \cos ^2\theta_{3} \sin ^2\theta_{1}
                        -  \frac 1 2\cos \delta  \cos \theta_{2} \sin 2   \theta_{3}\sin 2\theta_{1},   \\
   &      P^{1  2}_{2}&&=  &&       \sin 2 \theta_{2} \sin ^2\theta_{3},   
                              \\
   & P^{22}_{2}&&= &&\sin 2 \theta_{2} \cos ^2\theta_{3},   \\
     &      Q^{21}_{2}&&= &&
                        \sin 2\theta_{2} \sin ^2\theta_{3}\cos ^2\theta_{1} 
                        +\frac 1 2 \cos \delta  \cos \theta_{2} \sin 2 \theta_{3}   \sin 2\theta_{1}, \\
      &   Q^{31}_{2}&&= &&    \sin 2\theta_{2} \sin ^2\theta_{3}\sin^2 \theta_{1}
                        -\frac 1 2 \cos \delta \cos \theta_{2} \sin 2 \theta_{3} \sin 2\theta_{1} , \\
     & Q^{22}_{2}&&= &&\sin 2\theta_{2}   \cos^2\theta_{3}  \cos ^2\theta_{1}
                  -\frac 1 2 \cos \delta   \cos \theta_{2}\sin 2 \theta_{3} \sin 2\theta_{1},\\
      &    Q^{32}_{2}&&=&& \sin 2\theta_{2}  \cos ^2\theta_{3}\sin ^2\theta_{1}
                + \frac 1 2 \cos \delta  \cos \theta_{2} \sin 2 \theta_{3} \sin 2 \theta_{1}. \\[1em]
                 &    P^{1  1}_{3}       &&
                 =&& \sin 2\theta_{3}     \sin ^2\theta_{1} \sin^2 \theta_{2}
                       -\cos \delta \cos^2 \theta_{3} \sin 2\theta_{1}\sin \theta_{2},\\
        &      P^{21}_{3}    && = -&&  \sin 2\theta_{3}\sin^2\theta_{1} \sin^2 \theta_{2}
                                       -\cos \delta \sin^2 \theta_{3} \sin 2\theta_{1} \sin \theta_{2} , \\
         &     P^{31}_{3} &&=&& \sin 2 \theta_{3}
                            (\sin^2\theta_{1} \sin^2\theta_{2}-\cos^2\theta_{1} )
                            - \cos \delta   \cos 2 \theta_{3} \sin 2\theta_{1}\sin \theta_{2} ,\\
          &   P^{1  2}_{3}&&= -&&\sin 2 \theta_{3}\sin ^2\theta_{2}, \\
                  &   P^{22}_{3}&&= &&\sin 2 \theta_{3}\sin ^2\theta_{2}, \\
       &       P^{32}_{3}    &&=&&\sin 2 \theta_{3}  \cos ^2\theta_{2},\\
        &      Q^{21}_{3}      &&= -&&  \sin 2 \theta_{3}\sin^2 \theta_{1} 
                                         - \cos \delta \sin ^2\theta_{3} \sin2 \theta_{1} \sin \theta_{2}, \\
       &  Q^{31}_{3} &&=
                              -  &&\sin 2 \theta_{3}\cos^2 \theta_{1}
                            +\cos \delta\sin ^2\theta_{3}  \sin 2\theta_{1} \sin \theta_{2},\\
         &     Q^{22}_{3}   &&=&&  \sin 2 \theta_{3} \sin^2 \theta_{1}
                                         - \cos \delta   \cos ^2\theta_{3} \sin 2\theta_{1} \sin \theta_{2},\\
       &       Q^{32}_{3} &&=&&\sin 2 \theta_{3}\cos ^2\theta_{1}+ \cos \delta  \cos ^2\theta_{3} \sin 2\theta_{1}\sin \theta_{2}.\\[1em]
            \end{alignat*}
            \newpage
         %
                    \begin{alignat*}{3}
                           \kern -50pt       
                             &      P_\delta^{1  1} && = && \frac{   \cot \theta_{3}\tan \theta_{1}}   { \sin \theta_{2}}
      \Bigl[ \cos 2 \theta_{3} \left(1-\sin^2 \theta_{1}\cos^2 \theta_{2} \right)
                                                  +       (3\cos^2\theta_{1}\sin^2\theta_{2} - \cos^2\theta_{1} - \sin^2\theta_{2})\Bigr]            , \\
                             &      P_\delta^{21} && = &&
       \frac{  \tan \theta_{3}  \tan \theta_{1}}   { \sin \theta_{2}}
        \Bigl[
            \cos 2 \theta_{3} \left( 1-\sin^2 \theta_{1}\cos^2 \theta_{2}  \right)
           -(3\cos^2\theta_{1}\sin^2\theta_{2} - \cos^2\theta_{1} - \sin^2\theta_{2})
                                                            \Bigr]          , \\
                             &      P_\delta^{1  2} && = -&&   2 \sin 2 \theta_{3} \cot 2\theta_{1} \sin \theta_{2}, \\
                             &      P_\delta^{22} && = && 2  \sin 2 \theta_{3} \cot 2\theta_{1}\sin \theta_{2},\\
                 &     P_\delta^{31} && = &&    \frac{2 \sin 2\theta_{1} \sin \theta_{2} }{\sin 2\theta_{3} }, \\
                                              &      Q_\delta^{11} && = -&& \frac{2   \sin 2 \theta_{3} \sin \theta_{2}}
                                                                            { \sin 2\theta_{1}},  \\
         &    Q_\delta^{21} && =- &&   \frac{   \tan \theta_{3}\tan \theta_{1} } {\sin \theta_{2}}
                     \Bigl[\cos 2 \theta_{3}
                   \left(\cos^2\theta_{1}\cos ^2\theta_{2} -\sin ^2\theta_{2}\right)
             -\sin ^2\theta_{1} \sin ^2\theta_{2}+\cos ^2\theta_{1}\Bigr],          \\
        &  Q_\delta^{31}&&= && \frac{ \tan \theta_{3}\cot \theta_{1} }   {\sin \theta_{2}}
                               \Bigl[\cos 2 \theta_{3}
                               \left(\sin ^2\theta_{1} \cos ^2\theta_{2}-\sin ^2\theta_{2}\right)
                               -\cos ^2\theta_{1} \sin ^2\theta_{2}+\sin ^2\theta_{1}\Bigr],\\
   &     Q_\delta^{12} &&= &    &  \frac{2 \sin 2 \theta_{3}\sin \theta_{2} }{ \sin 2\theta_{1}},\\
         &   Q_\delta^{22} && = -&&    \frac{  \cot \theta_{3} \tan \theta_{1}} {\sin \theta_{2}}
             \Bigl[\cos 2 \theta_{3}
                                 \left(\cos^2\theta_{1}\cos ^2\theta_{2}    -\sin ^2\theta_{2}\right)
             +\sin ^2\theta_{1} \sin ^2\theta_{2}-\cos ^2\theta_{1}\Bigr],  \\
     &  Q_\delta^{32}&&= &&       \frac{  \cot\theta_{3}\cot \theta_{1}}     { \sin \theta_{2}}
                                                    \Bigl[\cos 2 \theta_{3}
                 \left(\sin ^2\theta_{1}\cos ^2\theta_{2}-\sin ^2\theta_{2}\right)
             +\cos ^2\theta_{1}\sin ^2\theta_{2}-\sin ^2\theta_{1}\Bigr]
                \end{alignat*}
      %
      (all other $P$s and $Q$s  vanish).

Observe that
\begin{alignat*}{3}
     \sum_{k=1}^3 \sum_{h=1}^2  P_{1}^{kh}  &=   \sin2\theta_{1},
 &&  \sum_{k=1}^3 \sum_{h=1}^2  Q_{1}^{kh}&&=
 \sin2\theta_{1}\cos^2\theta_{2},\\[1em]
  \sum_{k=1}^3 \sum_{h=1}^2  P_{2}^{kh}           &=
                                                     \sin 2\theta_{2}  \cos^2 \theta_{1},
                   &&        \sum_{k=1}^3 \sum_{h=1}^2  Q_{2}^{kh}&&=     \sin 2\theta_{2} ,
 \\[1em]
     \sum_{k=1}^3 \sum_{h=1}^2  P_{3}^{kh} & =
     \sin 2 \theta_{3}(\sin^2\theta_{1}   - \cos^2 \theta_{1}  \sin^2\theta_{2}   )
                                                  - 2\cos \delta   \sin 2 \theta_{1}\sin \theta_{2} \cos^2\theta_{3},\kern -170 pt&&&&\\[1em]
   \sum_{k=1}^3 \sum_{h=1}^2  Q_{3}^{kh} & =
                                            0,&& &&\\[1em]
    \sum_{k=1}^3 \sum_{h=1}^2  P_{\delta}^{kh} &= 2  
                                                 \sin 2 \theta_{1} \sin \theta_{2} \cot  \theta_{3} ,
                                                    && \sum_{k=1}^3 \sum_{h=1}^2  Q_{\delta}^{kh} &&= 0.
\end{alignat*}

We also note that $\dot\theta_{2}=0$ when
 $\theta_{2}=\frac{\pi}{2}$, so $\theta_{2}=\frac{\pi}{2}$ is an invariant hypersurface of the RG flow in $(\theta_{1},\theta_{2},\theta_{3},\delta)$-space.


\section{Operator mixing\label{app:lambda}}

At a global  fixed point $y_\alpha^*$, $\vartheta_\mu^*$, where $\beta_\alpha^*=\beta_\mu^*=0$, the operator mixing matrix is
\beq \Gamma_* =\left.
\bpm \frac{\partial \beta_\alpha}{\partial y_\beta} &  \frac{\partial \beta_\alpha}{\partial \vartheta_\nu} \\  
      \frac{\partial \beta_\mu}{\partial y_\beta} &  \frac{\partial \beta_\mu}{\partial \vartheta_\nu}\epm \right|_*
\eeq

This appendix gives the eigenvalues of the operator mixing matrix $\gamma_*$ in \eqref{eq:gamma*} at each of the six  fixed points at 1-loop.
We first prove that, at 1-loop,  $\Gamma_*$ is block diagonal,
\beq 
\Gamma_* = \left.\bpm \frac{\partial \beta_\alpha}{\partial y_\beta} &  0 \\  
      0 & \frac{\partial \beta_\mu}{\partial \vartheta_\nu} \epm \right|_{\vartheta=\vartheta*}
\eeq
for $\vartheta^*_\mu$ and any $y_\alpha$.

\begin{itemize}
\item 

As noted at the end of the section \ref{sec:CKM-PMNS}, at the fixed points $\vartheta_\mu^*$, all the $P$s and $Q$s must vanish, regardless of the prefactors involving $z_j$ and $z_j'$ in \eqref{eq:theta-dot} and \eqref{eq:delta-dot}. 
 This implies  that $\left.\frac{\partial \beta_\mu}{\partial y_\beta} \right|_*=0$.
\item
Next note that, in  equation \eqref{eq:A}, $A(\vartheta_\mu)$ is chosen to cancel the off-diagonal
components of 
$\frac{3}{2}( \Lambda V \Lambda' V^\dagger +  \Lambda' V^\dagger \Lambda) +i[A,\Lambda]$ for all $\theta_\mu$,
so the off-diagonal components of
\[\frac{\partial \Bigl\{  \frac{3}{2}( \Lambda V \Lambda' V^\dagger +  \Lambda' V^\dagger \Lambda) +i[A,\Lambda]\Bigr\}}{\partial \vartheta_\mu}\]
vanish for all values of $\vartheta_\mu$ by construction --- we only need to prove that the diagonal entries of the six matrices 
\[\frac{\partial ( \Lambda V \Lambda' V^\dagger +  \Lambda' V^\dagger \Lambda)}{\partial \vartheta_\mu}\]
vanish  at the six points $\vartheta^*_\mu$, and this is easily checked directly.  Exactly the same arguments applies to \eqref{eq:A'}.
\end{itemize}

On the $\theta_{2}=0$ hypersurface it is straightforward to calculate the eigenvalues of $\gamma$ at the fixed points: $\gamma_*$ in \eqref{eq:gamma*} is diagonal  in $(\theta_{1},\theta_{2}, \theta_{3},J)$ coordinates at the four fixed points
1, 2, 3 and 4 and the eigenvalues are just the four diagonal elements. 

For the two fixed points on the $\theta_{2}=\frac{\pi}{2}$ hypersurfaces the situation is rather more subtle.
First rotate the  $\theta_{1}$-$\theta_{3}$ plane is through $45^\circ$, to $\theta_\pm=\theta_{3}\pm\theta_{1}$, then at the fixed points 5 and 6, $\gamma_*$ diagonal.  However, expressing $\frac{\partial \dot J}{\partial J}$ in terms of $\delta$, the four choices
$(\theta_{1}^*,\theta_{2}^*,\theta_{3}^*,\delta^*)=(0,\frac{\pi}{2},\frac{\pi}{2},0)$, $(0,\frac{\pi}{2},\frac{\pi}{2},\pi)$,
$(\frac{\pi}{2},\frac{\pi}{2},0,0)$ and $(\frac{\pi}{2},\frac{\pi}{2},0,\pi)$
give the same physical mixing matrix $ V_{*,5}$,  but with  different $\gamma_*$ for the fixed point 5;  similarly the four choices
$(\theta_{1}^*,\theta_{2}^*,\theta_{3}^*,\delta^*)=(0,\frac{\pi}{2},0,0)$, $(0,\frac{\pi}{2},0,\pi)$,
$(\frac{\pi}{2},\frac{\pi}{2},\frac{\pi}{2},0)$ and $(\frac{\pi}{2},\frac{\pi}{2},\frac{\pi}{2},\pi)$,
 give the same physical mixing matrix $ V_{*,6}$  but with  different $\gamma_*$ for the fixed point 6: the differences lie in the fact that the eigenvalues are permuted, and some signs are changed.
This is due to the singularity when $\theta_{2}=\frac{\pi}{2}$ and $\delta=0$ or $\pi$ described in section \ref{sec:CKM-PMNS}: the space of physical mixing matrices when  $\theta_{2}=\frac{\pi}{2}$, $\delta=0,\pi$ is only 1-dimensional, not 2-dimensional.  

To unravel this consider $\gamma_*$ in $(\theta_+, \theta_-,\theta_{2})$ coordinates on the hypersurface  $\theta_{2}=\frac{\pi}{2}$.  
 In the left-hand diagram in figure \ref{fig:SU3},
$\theta_+=\frac{\pi}{2}$  at the two vertices corresponding to fixed point 5 (purple), 
but $\theta_-$ is different at these two vertices, it is not a good coordinate there: 
similarly $\theta_-=0$  at the two vertices corresponding to fixed point 6 (green),
but $\theta_+$  is different at these two vertices.  
We therefore use $\theta_+$ for fixed point 5 and $\theta_-$ for fixed point 6. 

For convenience we define
\beq
\zeta_{j k} = \frac{z_k}{z'_j(1+z_k)}\qquad \mbox{and} \qquad
 \zeta'_{j k} = \frac{z'_k}{z_j(1+z'_k)},\eeq
as these combinations appear frequently and this notation tidies up the following formulae.

From \eqref{eq:theta-dot} and appendix \ref{app:PQ}:
\begin{align*}
  \beta_+&= \frac{ d \theta_+}{d t}=\frac{3}{32\pi^2}
                 \left[ y_3^{\prime 2}\left( \zeta'_{11} -\zeta'_{12}\right)
                 -y_3^2 \zeta_{31}\right]
                 \sin 2\theta_+\quad  \mbox{for} \quad \theta_2=\frac{\pi}{2},\ \delta=0;\\
  \beta_-&= \frac{ d \theta_-}{d t}=\frac{3}{32\pi^2}
                 \left[y_3^{\prime 2}\left( \zeta'_{11} -\zeta'_{12}\right)
                 -y_3^2 \zeta_{31}\right]
                                 \sin 2\theta_-\quad \mbox{for} \quad \theta_2=\frac{\pi}{2},\ \delta=\pi,        
\end{align*}
and
\begin{align*}
  \frac{\partial \beta_+}{\partial \theta_+}&= \frac{3}{16\pi^2}
   \left[y_3^{\prime 2}\left( \zeta'_{11} -\zeta'_{12}\right)  -y_3^2 \zeta_{31}  \right] \cos 2\theta_+\qquad \mbox{for} \quad\theta_2=\frac{\pi}{2},\  \delta=0;\\
  \frac{\partial \beta_-}{\partial \theta_-}&= \frac{3}{16\pi^2}
   \left[y_3^{\prime 2}\left(  \zeta'_{11} -\zeta'_{12}\right)  -y_3^2 \zeta_{31} \right] \cos 2\theta_-\qquad \mbox{for} \quad\theta_2=\frac{\pi}{2},\  \delta=\pi.      
\end{align*}

Using $\theta_+=\frac{\pi}{2}$,  $\delta=0$ for fixed point 5, and $\theta_-=0$,
$\delta=\pi$ 
for fixed point 6 yields
\begin{align*}
  \lambda_+=
  \left.\frac{\partial \beta_+}{\partial \theta_+}\right|_{\theta_+=\frac{\pi}{2},\, \delta=0}
  &  = \hskip 8pt  \frac{3}{16\pi^2}
           \left[y_3^{\prime 2}\left( \zeta'_{12} -\zeta'_{11}\right)  +y_3^2 \zeta_{31}  \right]
                                                                                                    \quad \mbox{at fixed point 5};\\
  \lambda_-=\left.\frac{\partial \beta_-}{\partial \theta_-}\right|_{\theta_-=0,\,\delta=\pi}
       &= -\frac{3}{16\pi^2}
       \left[y_3^{\prime 2}\left(  \zeta'_{12} -\zeta'_{11}\right)  +y_3^2 \zeta_{31} \right]
                                                                                        \quad \mbox{at fixed point 6.}
\end{align*}

For  $\lambda_2=\left. \frac{\partial \beta_2}{\partial \theta_2}\right|_*$
 on the $\theta_{2}=\frac{\pi}{2}$ hypersurface  $\theta_*=(0,\frac{\pi}{2},\frac{\pi}{2})$ and $\theta_*=(\frac{\pi}{2},\frac{\pi}{2},0)$ give different eigenvalues for $\lambda_2$ at fixed point 5, and  $\theta_*=(0,\frac{\pi}{2},0)$ and $\theta_*=(\frac{\pi}{2},\frac{\pi}{2},\frac{\pi}{2})$ give different eigenvalues for $\lambda_2$ at fixed point 6. The reason for this can be again understood  from figure \ref{fig:SU3}:
for fixed point 5 one eigenvalue is from the direction of the 2-5 edge and the other from the direction of the 4-5 edge; for fixed point 6 one eigenvalue is from the direction of the 1-6 edge and the other from the direction of the 
3-6 edge: these are labelled accordingly below.

The four eigenvalues of $\gamma_*$ at each of the fixed points 1-6 obtained from this strategy are:
{\scriptsize{\begin{alignat*}{3}
{\bm 1)} \quad     {\scriptstyle{ (\theta^*_{1},}}&   {\scriptstyle{\theta^*_{2},\theta^*_{3},J^*)=(0,0,0,0)}}  
&
     {\bm 2)} \quad    {\scriptstyle{(\theta^*_{1},}}&
       {\scriptstyle{\theta^*_{2},\theta^*_{3},  J^*)=\bigl(0,0,\frac{\pi}{2},0\bigr) }} \\[0.5em]
  \lambda_{{\mathbf 1},1}&= \hskip 8pt \frac{3 }{16\pi^2}\left(
               y_3^2\zeta_{11}
                +     y_3^{\prime\, 2}\zeta_{11}'
                           \right),& 
 \lambda_{{\mathbf 2},1} & =\frac{3}{16\pi^2}
                 \left(y_3^2\zeta_{21} + y_3^{\prime\, 2} \zeta_{12}'
                            \right),  \\
  \lambda_{{\mathbf 1},2}&= \hskip 8pt \frac{3 }{16\pi^2}
                \left(
               y_3^2\zeta_{22}+y_3^{\prime\, 2}\zeta_{22}'  
                           \right),&
    \lambda_{{\mathbf 2},2} & =\frac{3}{16\pi^2}\left(
                y_3^2 \zeta_{12} + y_3^{\prime\, 2}\zeta'_{21}  
                             \right),\\
  \lambda_{{\mathbf 1},3}&=-\frac{3 }{16\pi^2}  \left[
               y_3^2  \left(\zeta_{31} -\zeta_{32}
                            \right)
                +y_3^{\prime\, 2}\left(\zeta'_{31} - \zeta'_{32}
                \right)
                           \right],  \qquad &
   \lambda_{{\mathbf 2},3} & =\frac{3}{16\pi^2}
                 \left[
                y_3^2 \left(\zeta_{31} -\zeta_{32}
                               \right)
                 +y_3^{\prime\, 2}\left(\zeta_{31}' -\zeta_{32}'
            \right)
                 \right] =-  \lambda_{{\mathbf 1},3},    \\
  \lambda_{{\mathbf 1},J}&
            =\lambda_{{\mathbf 1},1}+\lambda_{{\mathbf 1},2}+\lambda_{{\mathbf 1},3}\; ;
                         &
  \lambda_{{\mathbf 2},J}
     &=\lambda_{{\mathbf 2},1}+\lambda_{{\mathbf 2},2}+\lambda_{{\mathbf 2},3}\; ;
       \\[2.5em]
%
 {\bm 3)} \quad        {\scriptstyle{(\theta^*_{1},}}
  &   {\scriptstyle{\theta^*_{2},\theta^*_{3},J^*)=\bigl(\frac{\pi}{2},0,\frac{\pi}{2},0\bigr) }} &
        {\bm 4)} \quad      {\scriptstyle{ (\theta^*_{1},}}&   {\scriptstyle{\theta^*_{2},\theta^*_{3},  J^*)=\bigl(\frac{\pi}{2},0,0,0\bigr)}}    \\[0.5em]
                \lambda_{{\mathbf 3},1} & =-\frac{3}{16\pi^2}\left(
       y_3^2\zeta_{21}    + y_3^{\prime\, 2}\zeta'_{12} \right)= -\lambda_{{\mathbf 2},1},
 &
         \lambda_{{\mathbf 4},1} & = -\frac{3}{16\pi^2}\left(
                y_3^2 \zeta_{11} +  y_3^{\prime\, 2}\zeta'_{11} 
                            \right) = -\lambda_{{\mathbf 1},1},\\
                        \lambda_{{\mathbf 3},2} & =-\frac{3}{16\pi^2}\left[
                                y_3^2 \left(
          \zeta_{11} -    \zeta_{12}   
                                     \right)
                          -     y_3^{\prime\, 2} \zeta'_{31}     \right],&
                      \lambda_{{\mathbf 4},2} & = \hskip 8pt \frac{3}{16\pi^2}\left[
              y_3^2(\zeta_{22}-\zeta_{21})+ y_3^{\prime\, 2}\zeta'_{32} 
                           \right], \\
   \lambda_{{\mathbf 3},3} & = -\frac{3}{16\pi^2} \left[
                                 y_3^2\zeta_{32}+   y_3^{\prime\, 2} \left( 
                             \zeta'_{22}-\zeta'_{21}\right)\right]= -\lambda_{{\mathbf 4},3}, &
              \lambda_{{\mathbf 4},3} & = \hskip 8pt\frac{3 }{16\pi^2}
                 \left[y_3^2\zeta_{32} 
                 +y_3^{\prime\, 2} (\zeta'_{22}-\zeta'_{21}    )
                                              \right],\\
  \lambda_{{\mathbf 3},J} &
  =    \lambda_{{\mathbf 3},1}+\lambda_{{\mathbf 3},2}+\lambda_{{\mathbf 3},3}\; ;&
            \lambda_{{\mathbf 4},J} & 
            =\lambda_{{\mathbf 4},1}+\lambda_{{\mathbf 4},2}+\lambda_{{\mathbf 4},3}\; ;                 \\[2.5em]
  %
%
  {\bm 5)} \quad       {\scriptstyle{(\theta^*_{1},}}&
                   {\scriptstyle{\theta^*_{2},\theta^*_{3}, J^*)
   =\bigl(0,\frac{\pi}{2},\frac{\pi}{2},0\bigr)\quad\mbox{or}\quad  \bigl(\frac{\pi}{2}, \frac{\pi}{2},0,0\bigr)}}&      {\bm 6)} \quad       {\scriptstyle{(\theta^*_{1},}}
           &   {\scriptstyle{\theta^*_{2},\theta^*_{3},J^*)=
    \bigl(0,\frac{\pi}{2},0,0\bigr)\quad \mbox{or}\quad \bigl(\frac{\pi}{2},\frac{\pi}{2},\frac{\pi}{2},0\bigr)}} \\[0.5em]
    \lambda_{{\mathbf 5},+} & =\hskip 6pt \frac{3}{16\pi^2}\left[
               y_3^2 \zeta_{31}
                -y_3^{\prime\, 2}\left(
           \zeta'_{11} -  \zeta'_{12}
                \right)
                \right],&   \lambda_{{\mathbf 6},-} & =-\frac{3}{16\pi^2}
                \left[
                y_3^2\zeta_{31}-   y_3^{\prime\, 2}\left(
            \zeta'_{11}
                -\zeta'_{12}
                \right)
                           \right]= -\lambda_{{\mathbf 5},+},\\
   \lambda_{{\mathbf 5},25} & = \hskip 6pt \frac{3}{16\pi^2}
                   \left[
                   y_3^2\left(
             \zeta_{21}  - \zeta_{22}
                   \right)
                   -y_3^{\prime\, 2} \zeta'_{32}
                   \right]= -\lambda_{{\mathbf 4},2},&
       \lambda_{{\mathbf 6},16} & =-\frac{3}{16\pi^2}
              \left[  y_3^2   \zeta_{22}
                   +  y_3^{\prime\, 2}\zeta'_{22}
                   \right] = -\lambda_{{\mathbf 1},2},\\
  \lambda_{{\mathbf 5},45}&        =       
                   -\frac{3}{16\pi^2}\left(
             y_3^2   \zeta_{12}
              +y_3^{\prime\, 2} \zeta'_{21}
                          \right) = -\lambda_{{\mathbf 2},2},
                             &   \lambda_{{\mathbf 6},36}&= 
                \hskip  8pt   \frac{3}{16\pi^2}\left[
                  y_3^2\left(
          \zeta_{11}
                  - \zeta_{12}
                  \right)
                  -y_3^{\prime\, 2}\zeta'_{31}
                              \right]= -\lambda_{{\mathbf 3},2},\\
  \lambda_{{\mathbf 5},J}&
                           =    \lambda_{{\mathbf 5},+}+\lambda_{{\mathbf 5},25}+\lambda_{{\mathbf 5},45}\; ;&
         \lambda_{{\mathbf 6},J} &
    =    \lambda_{{\mathbf 6},-}+\lambda_{{\mathbf 6},16}+\lambda_{{\mathbf 6},36}.
    %
  %
 %
         \end{alignat*}
       }}
       
       \medskip
     \noindent     Note that,
for all six fixed points $A=1,\ldots,6$, $\lambda_{A,J}$ is the sum of the other three eigenvalues, 
so
\[  Tr(\gamma_{*,A})=2 \lambda_{A,J}; \]
 also the six traces can be put into two  subsets $\{1,3,5\}$  and $\{2,4,6\}$, satisfying 
\[Tr(\gamma_{*,1})+Tr(\gamma_{*,3})+ Tr(\gamma_{*,5})
=Tr(\gamma_{*,2})+Tr(\gamma_{*,4})+ Tr(\gamma_{*,6})=0,\]
corresponding to even and odd permutations of the permutation group $S_3$,
$\{V_1,V_3, V_5\}$ and $\{V_2,V_4, V_6\}$  in equations \eqref{eq:S_3}.

\end{document}